\documentclass[usenatbib]{mn2e}
\usepackage{amsmath}
\usepackage{graphicx}
\usepackage{pgfplots}
\usepackage{listings}
\pgfplotsset{compat=1.5}

\def\lesssim{\mathrel{\hbox{\rlap{\hbox{\lower4pt\hbox{$\sim$}}}\hbox{$<$}}}}
\def\gtrsim{\mathrel{\hbox{\rlap{\hbox{\lower4pt\hbox{$\sim$}}}\hbox{$>$}}}}

\def\arcsec{\hbox{$^{\prime\prime}$}}

\def\farcs{\hbox{$.\!\!^{\prime\prime}$}}

\newcommand{\mamo}[1]{\mbox{$#1$}}
\def\phs{\phantom{\mamo{-}}}     % Phantom minus sign for aligning columns in tables
\newcommand{\unit}[1]{\ifmmode \:\mbox{\rm #1}\else \mbox{#1}\fi}
\newcommand{\sbr}[1]{_{\rm #1}}

\newcommand{\px}{\unit{px}}
\newcommand{\chisq}{\mamo{\chi^2}}
\newcommand{\Chisq}{\mamo{X^2}}

\newcommand{\Seclabel}[1]{\label{sec:#1}}

\newcommand{\Applabel}[1]{\label{sec:#1}}
\newcommand{\Eqlabel}[1]{\label{eq:#1}}
\newcommand{\Figlabel}[1]{\label{fig:#1}}
\newcommand{\Tablabel}[1]{\label{tab:#1}}
\newcommand{\Secref}[1]{Section~\ref{sec:#1}}

\newcommand{\Appref}[1]{Appendix~\ref{sec:#1}}
\newcommand{\Eqref}[1]{Equation~(\ref{eq:#1})}
\newcommand{\Figref}[1]{Fig.~\ref{fig:#1}}
\newcommand{\Tabref}[1]{Table~\ref{tab:#1}}

\definecolor{dkgreen}{rgb}{0,0.6,0}
\definecolor{gray}{rgb}{0.5,0.5,0.5}
\definecolor{mauve}{rgb}{0.58,0,0.82}

\title[PSF Model Validation]{Validation of PSF Models for HST and Other Space-Based Observations}
\author[B.R. Gillis {\it et al.}]{Bryan R. Gillis$^{1}$\thanks{E-mail: b.gillis@roe.ac.uk}, Tim Schrabback$^{2}$, Ole Marggraf$^{2}$, 
\newauthor Rachel Mandelbaum$^{3}$, Richard Massey$^{4,5}$, Jason Rhodes$^{6,7}$, Andy Taylor$^{1}$ \\
$^{1}$Institute for Astronomy, University of Edinburgh, Royal Observatory Edinburgh, Edinburgh, EH9 3HJ, United Kingdom. \\
$^{2}$Argelander Institute for Astronomy, University of Bonn, Auf dem H\"ugel 71, 53121 Bonn, Germany \\
$^{3}$McWilliams Center for Cosmology, Department of Physics, Carnegie Mellon University, Pittsburgh, PA 15213, USA, \\
$^{4}$Centre for Extragalactic Astronomy, Department of Physics, Durham University, Durham DH1 3LE, U.K., \\
$^{5}$Institute for Computational Cosmology, Durham University, South Road, Durham DH1 3LE, U.K, \\
$^{6}$Jet Propulsion Laboratory, California Institute of Technology, 4800 Oak Grove Dr., Pasadena, CA 91109, USA, \\
$^{7}$California Institute of Technology, 1200 E. California Blvd., CA 91125, USA}

\begin{document}

\date{17 Jun 2019}

\pagerange{\pageref{firstpage}--\pageref{lastpage}} \pubyear{2019}

\maketitle

\label{firstpage}

\begin{abstract}

\noindent Forthcoming space-based observations will require high-quality point-spread function (PSF) models for weak gravitational lensing measurements. One approach to generating these models is using a wavefront model based on the known telescope optics. We present an empirical framework for validating such models to confirm that they match the actual PSF to within requirements by comparing the models to the observed light distributions of isolated stars. We apply this framework to Tiny Tim, the standard tool for generating model PSFs for the \textit{Hubble Space Telescope} (HST), testing its models against images taken by HST's Advanced Camera for Surveys in the Wide Field Channel. We show that Tiny Tim's models, in the default configuration, differ significantly from the observed PSFs, most notably in their sizes. We find that the quality of Tiny Tim PSFs can be improved through fitting the full set of Zernike polynomial coefficients which characterise the optics, to the point where the practical significance of the difference between model and observed PSFs is negligible for most use cases, resulting in additive and multiplicative biases both of order $\sim 4\times10^{-4}$. We also show that most of this improvement can be retained through using an updated set of Zernike coefficients, which we provide.

\end{abstract}


\section{Introduction}

Optical observations are fundamentally limited by the point-spread function (PSF) of the observing instrument and conditions. This describes the manner in which a point light source will be affected to spread the paths of photons, displaying not as a point but as an extended profile. Various effects contribute to this, including atmospheric diffraction, diffraction due to obscurations within the instrument, aberrations in the instruments optics, and various detector effects \citep{KriHooSto11}. Accurate knowledge of the PSF is thus necessary for any pursuit which requires knowledge of the undistorted light distribution of an object.

For instance, weak gravitational lensing uses the shapes of objects before PSF distortion as an estimator for the gravitational shear field. Elliptical PSFs will contribute to the observed ellipticities of sources, and since in general the PSF spreads detected photons, this contributes to circularizing them. These effects respectively result in additive and multiplicative biases in the ellipticities of sources, and so it is necessary to have a proper characterization of the PSF in order to make unbiased measurements of source shapes \citep{KaiSquBro95,LupKai97,HoeFraKui98,BarSch99,Kai00}.

For ground-based observations, the PSF is dominated by atmospheric conditions which vary on short timescales, and so its profile in any given observation must be determined empirically, using point-source objects such as stars to sample it and interpolating appropriately \citep{JarJai04,Row10,HamMiyOku13,LuZhaDon17}. For space-based observations, however, it is possible to design a system so that the PSF is relatively stable, determined only by instrumental effects such as diffraction due to obscurations, optical aberrations, and polishing errors. It is therefore theoretically possible to instead model the PSF based on an understanding of the instrument's optics \citep{KriHooSto11}. This is particularly useful for measurements which require high precision such as the upcoming Euclid mission \citep{LauAmiArd11}, as for an empirical model, the number of observed stars might not be sufficient to overcome noise, particularly in the wings of the PSF. Additionally, the Euclid mission is designed to have as stable a PSF as possible, making it an ideal candidate for using a model PSF.

Before such PSF models can be used for measurements, it will be necessary to validate them through comparisons with observations, to ensure that the models actually are faithful recreations of the instrumental PSF. It is not immediately obvious how best to perform such validation, and it is necessary to weight the chances and impacts of false positives and false negatives for any validation procedure. In this paper, we discuss the decisions that must be made, and we present a validation framework which is targeted at ensuring a PSF model is suitable for weak lensing measurements.

We demonstrate this framework by applying it to Tiny Tim \citep{Kri93,KriHooSto11} model PSFs and comparing them to observations taken with the \textit{Hubble Space Telescope} (hereafter ``HST''). Tiny Tim has a long history of use in processing HST images, for purposes such as e.g. measuring the sizes of star clusters \citep{WhiZhaLei99} and distant galaxies \citep{vanFraKri08}, measuring the width of a protostar jet \citep{BurStaWat96}, strong gravitational lensing observations \citep{SurBla03}, and weak gravitational lensing observations \citep{LeaMasKne07}.

The literature is sparse, however, when it comes to validation tests of Tiny Tim. Of note, \citet{vanBelH12} tested Tiny-Tim-based PSF models against a sample of 46 isolated stars and found a $4$ per cent discrepancy in the amount of light outside of an $0\farcs2$ circular aperture. Imperfections in the model were also noted in \citep{KriHooSto11}, where they note ``For subtraction in the wings, usually an observed reference PSF provides superior results than using a model, as the match to the fine scale structure is better.'' \citet{HofAnd17} also compared Tiny Tim to an empirical PSF, and found that the empirical model outperformed Tiny Tim by a factor of 2 in their quality-of-fit metric. Insight can also be gained from comparisons of observed PSFs with models generated in other manners, such as \citet{JeeBlaSir07}'s analysis of PSF models generated with a principle-component analysis technique. 

Further testing of Tiny Tim is thus warranted to see if these results can be replicated with a larger sample and to test further properties of the PSF beyond its light profile which might affect lensing measurements and other applications which rely on Tiny Tim models.

In \Secref{Data_and_Models} of this paper, we discuss the data and models we use to test our framework. In \Secref{Testing_Procedure}, we present our proposed testing framework for the validation of PSF models, and in \Secref{Control_Tests} we present and discuss our control tests.We discuss our findings in \Secref{Results} and conclude in \Secref{Conclusions}.

\section{Data and Models}
\Seclabel{Data_and_Models}

\subsection{Data}
\Seclabel{Data}

For our tests, we use data from the HST observed with the Advanced Camera for Surveys \citep[``ACS'', ][]{AviGroAnd16} in the wide-field channel using the F606W filter. We use a set of 217 pointings observed between the dates July 9 2009 and July 16 2011, selected for high stellar density. These are generally observations of globular clusters such as NGC 104, but also include images of the nearby dwarf galaxy NGC 185. Each pointing comprises a set of two $4096 \times 2048$ pixel images, with an approximate pixel scale of $0\farcs05/\mathrm{pixel}$, for an observed area of $\sim 5.8$ arcmin$^2$ for each image. The pointings cover regions of varying stellar number density and have varying exposure, resulting in a greatly variable number of usable stars per image (from $\sim 10$ to $\sim 200$, with significant bimodality in the distribution).

The full reduction pipeline used for our data is detailed in \citet{SchAppDie18}, but in short, the images are reduced primarily with the standard ACS calibration pipeline \texttt{CALACS}. The images are corrected for the effects of charge-transfer inefficiency through the procedure presented in \citet{MasSchCor14}. Each of the images is then processed by the MultiDrizzle tool \citep{KoeFruHoo03} to correct for cosmic rays, but the final ``drizzle'' step (which stacks exposures together) is not performed, as this alters the effective PSF of the stacked image and complicates analysis of it. Since we work only with unstacked images from this point forward, we must be careful to only perform operations that are valid on unstacked images, which means e.g. avoiding interpolation.

\subsection{Tiny Tim PSF Models}
\Seclabel{PSF_models}

For this work, we test the PSF models generated by Tiny Tim \citep{Kri93,KriHooSto11}, the standard tool for generating model PSFs for the HST. Tiny Tim calculates the PSF through a wavefront model, using the known obscuration pattern of the HST and assuming that all obscuration happens in a single plane (see section 3.1 of \citealt{KriHooSto11} for justification of this assumption). The wavefront is represented by an $n \times n$ array where $n$ is sufficiently large that the wavefront is Nyquist sampled, and the wavefront is propagated to the focus with a fast Fourier transform. The square of the magnitude of the electromagnetic field is then taken to produce the model PSF. Tiny Tim also includes prescriptions to model aberrations due to defocus and polishing errors, and can also include a prescription for jitter if desired.

The effective PSF is also affected by the diffusion of charge between neighbouring pixels. This step can be modelled by Tiny Tim if desired, but it is often more useful for the user to request a subsampled model PSF. Such a model PSF can be shifted so that it represents the proper placement of the point source relative to the centre of the nearest pixel. The subsampled model PSF can then be rebinned and convolved with the $3 \times 3$ charge diffusion kernel provided by Tiny Tim to generate the final model PSF, which is what we do in our analysis.

We test Tiny Tim with a use-case similar to the approach of e.g. \citet{RhoMasAlb07}, who fit focus offset value input to Tiny Tim for each image by finding the value which resulted in PSF models being generated which best matched the observed stars. We use the downloadable Tiny Tim tool, version 7.5\footnote{\texttt{http://tinytim.stsci.edu/static/tinytim-7.5.tar.gz}}, to generate model PSFs. The tool's first executable, \texttt{tiny1}, asks the user a series of questions which allow it to determine the PSF to be generated, and it stores the necessary information in a parameter file. This file also contains values which the tool does not ask the user about but which can be modified, such as the assumed coma and astigmatism. As we expect users will not do this unless it is shown to be necessary, we will start by testing the scenario in which the only adjusted parameters are those which the tool asks about. These parameters and our selections for them are:

\begin{itemize}
	\item \textbf{Camera:} 15 (ACS - Wide Field Channel)
	\item \textbf{Detector:} Chosen per image based on the ``\texttt{CCDCHIP}'' header keyword
	\item \textbf{Position on Detector:} Chosen per star, taking the nearest of a fixed $32 \times 16$ grid of positions. This constrains the number of PSFs which need to be generated, and was found to not noticeably impact our results per our analysis on control fields (see \Secref{Control_Tests}).
	\item \textbf{Filter Passband:} F606W
	\item \textbf{Spectrum:} 1; 15 (Use the K7V spectrum, which is chosen to represent a typical star in the sample. We discuss the impact of this in \Secref{Control_Tests}.)
	\item \textbf{PSF Diameter:} 2.0 arcsec
	\item \textbf{Focus-Secondary Mirror Despace:} Fit per image. As HST ``breathes'' due to its passing in and out of the Earth's shadow, the position of the focus relative to the secondary mirror changes over time and cannot be perfectly predicted for any given observation \citep{Kri03,AndKin06}. We thus have to fit the best focus position by simulating multiple sets of PSFs for each image.
\end{itemize}

The chosen Position on Detector value is used by Tiny Tim to determine field dependent aberrations and charge diffusion, which are used in the generation of the PSF. If the user desires, it is possible to edit the generated parameter file to change the Zernike coefficients which determine aberrations, as we test in \Secref{Further_TT_Testing}. The values in this parameter file correspond to the values at the centre of the field, and Tiny Tim adds these to position-dependent offsets for each coefficient to determine the value to use for each generated PSF.

A finely-subsampled model PSF is then generated through running \texttt{tiny2} with the generated parameter file. For the ACS, it is also necessary to apply distortion due to the fact that it is installed off-axis. This is done through the \texttt{tiny3} command, which also allows the user to determine their desired subsampling factor for the final model PSF. We choose a subsampling factor of $8 \times$, based on our analysis on control fields (see \Secref{Control_Tests}). This factor is large enough to allow us to shift the model PSF to adequately match the proper subpixel centre for any given star. We apply this shift for each star, based on the relative positions of the best-fit centre of the star and the subsampled model PSF, and then rebin the model PSF to the same pixel scale as the detector. Finally, we apply the charge-diffusion kernel provided for each subsampled model PSF to generate the final model PSF for each star. At this stage, we compare the subpixel centre for the star and rebinned model PSF. If the difference is greater than the subsampled pixel scale, we return to the subsampled model and shift it an additional amount corresponding to this offset, and then rebin and convolve with the charge-diffusion kernel again. We repeat this process until the centre converges.

\section{PSF Validation Framework}
\Seclabel{Testing_Procedure}

\subsection{Designing Tests}
\Seclabel{designing_tests}

For a given PSF model, ideally its predictions of the PSF from a point source will differ from its observed light profile only in noise and the effects of environment (i.e. other nearby sources on the sky). We can therefore propose various statistical tests which we expect that an ideal model will typically pass. The most straightforward such test is a $\chisq$ test on the fluxes of a set of pixels, comparing the predicted flux from the model to the observed flux surrounding an isolated point source, using a proper characterisation of the noise and taking into account the number of parameters of the PSF model which must be fitted to the data. Such a test has the following issues, however:
\begin{enumerate}
	\item It tests the statistical significance of a departure from a perfect model, but a statistically significant failure of the test does not mean that the model is unusable. Depending on what the model is used for, its imperfections may have negligible or even no effect on the resulting measurements. 
	\item Truly isolated point sources are difficult to identify. Stars are effectively point sources, and the brighter stars can be reasonably distinguished from galaxies through a selection in size-magnitude space, but many apparent stars are in fact unresolved binary star systems (e.g. \citealt{Lad06} finds $\sim 31$ per cent of stellar systems host more than one star), and this binary nature will result in the light profile being larger and more elliptical than that of a point source.
	\item This test requires that the background behind each point source be uniformly zero. If the background is not perfectly subtracted, this will result in the $\chisq$ statistic for the comparison being spuriously large.
\end{enumerate}

We discuss these issues in the following subsections.

\subsubsection{Relevant Quantities}
\Seclabel{relevant_quantities}

As the primary motivation of this work is to develop a validation framework for PSF models to be used for gravitational lensing, we desire our tests to be failed if and only if imperfections in the PSF would lead to significant errors in the measured gravitational lensing signals. To determine what tests might be necessary, let us briefly provide an overview of how lensing signals are measured. Lensing signals are determined from the measured shear estimates derived from galaxies light distributions. If galaxies are assumed to be elliptical and randomly oriented, then their measured ellipticities\footnote{For ellipticity, we use the definition $\left|e\right| = (1-r)/(1+r)$, where $r$ is the axis ratio of the galaxy.} will provide unbiased estimates of the amount of shear in the absence of pixelisation and the influence of a PSF. The influence of the PSF will affect the apparent ellipticities of observed galaxies, and so it must be properly characterized for an accurate measurement of a lensing signal. See e.g. \citet{Man15} for further details on the theory and methods underlying weak gravitational lensing.

The two most common approaches to shear-measurement algorithms are moments-based and model-fitting approaches. Moments-based approaches primarily use measurements of the weighted quadrupole moments of observed galaxies and PSFs\footnote{Methods typically also use higher-order moments for corrections to their estimates. It is not necessary to take this into account here, as the quadrupole moments are the most significant factors going into the estimates, and they sufficiently constrain the PSF fitting.} to calculate the most likely ellipticity of the undistorted galaxy image. Model-fitting approaches use model profiles for the undistorted galaxies, convolve these with the PSFs, and test the convolved models against the observed galaxies to find the best-fitting models, and the ellipticities of these models are then used. Often, an approach called ``Metacalibration'' \citep{SheHuf17} is applied, testing the impact of perturbations to the data on the shear estimates and using this to calibrate the shear estimates. This still assumes a perfect PSF model in the most straightforward implementations though, and as such will not remove the sensitivity to imperfections in the model.

For the purposes of testing PSFs, the manner in which moments-based shear-measurement algorithms work is particularly illuminating. These methods are primarily sensitive to the quadrupole moments of the PSFs, and so PSF models which have imperfections in these moments will have direct effects on the estimates of galaxies' undistorted ellipticities. The PSF's dipole moments are also relevant, as they will affect the determination of its centre and thus the calculations of its quadrupole moments. As the PSF conserves flux, its monopole moment is only useful for normalizing other moments.

As used in gravitational lensing analysis, the normalised weighted multipole moments are defined as:
\begin{align}
	\Eqlabel{moment_definitions}
	M_{x} &= \frac{1}{m\sbr{0}}\iint x \; w(x,y) \; I(x,y) \; d^2 A\mathrm{,} \\ \nonumber
	M_{y} &= \frac{1}{m\sbr{0}}\iint y \; w(x,y) \; I(x,y) \; d^2 A\mathrm{,} \\ \nonumber
	M_{xx} &= \frac{1}{m\sbr{0}}\iint x^2 \; w(x,y) \; I(x,y) \; d^2 A\mathrm{,} \\ \nonumber
	M_{yy} &= \frac{1}{m\sbr{0}}\iint y^2 \; w(x,y) \; I(x,y) \; d^2 A\mathrm{,} \\ \nonumber
	M_{xy} &= \frac{1}{m\sbr{0}}\iint xy \; w(x,y) \; I(x,y) \; d^2 A\mathrm{,}
\end{align}
where
\begin{equation}
	m\sbr{0} = \iint w(x,y) \; I(x,y) \; d^2 A\mathrm{,} \\ \nonumber
\end{equation}
$w(x,y)$ is a weight function which quickly approaches zero at some finite radius, $d^2 A$ represents an integral over the full relevant area, and the central position $(x,y)=(0,0)$ is generally chosen so that $M_{x}$ and $M_{y}$ will be zero for some weight function. In the case of a pixelized image, the integral can be replaced with a summation. The two ellipticity components of an object can be calculated through one of many algorithms using these moments. For example, one of the simpler algorithms is:
\begin{align}
	\Eqlabel{ellipticity_from_moments}
	\hat{\epsilon}\sbr{1} &= \frac{M_{xx}-M_{yy}}{M_{xx}+M_{yy}+2\sqrt{\left|\mathbf{M}\right|}}\mathrm{,} \\ \nonumber
	\hat{\epsilon}\sbr{2} &= \frac{2M_{xy}}{M_{xx}+M_{yy}+2\sqrt{\left|\mathbf{M}\right|}}\mathrm{,} 
\end{align}
\citep{SeiSch97} where:
\begin{equation}
	\left|\mathbf{M}\right| = M_{xx}M_{yy} - M_{xy}^2 \mathrm{.}
\end{equation}
Note that this particular algorithm is not commonly used, as the reliance on $\left|\mathbf{M}\right|$, which includes the products of moments, gives the algorithm high sensitivity to noise. More commonly, algorithms instead use simply $M_{xx}+M_{yy}$ in the denominator and apply other corrections which do not rely on the products of moments \citep[e.g.][]{KaiSquBro95,LupKai97,HoeFraKui98}. However, the mathematical simplicity of this algorithm is more beneficial for our purpose here of estimating the impact of model PSF inaccuracies on shear estimates.

In the simplified scenario of unweighted moments, the normalised quadrupole moments of the undistorted galaxy can be calculated from the difference of the normalised moments of the observed galaxy and PSF:
\begin{equation}
	\Eqlabel{moment_subtraction}
	\mathbf{M}^{(u)} = \mathbf{M}^{(o)} - \mathbf{M}^{(p)}\mathrm{,}
\end{equation}
where $\mathbf{M}^{(u)}$ represents the quadrupole moments of the undistorted galaxy, $\mathbf{M}^{(o)}$ those of the observed galaxy, and $\mathbf{M}^{(p)}$ those of the PSF. Combined with \Eqref{ellipticity_from_moments}, we can identify and define the most relevant moments-based quantities for the PSF as:
\begin{align}
	\Eqlabel{relevant_quadrupole_terms}
	M\sbr{+} &= M_{xx}-M_{yy} \mathrm{,} \\ \nonumber
	M\sbr{\times} &= 2M_{xy} \mathrm{,} \\ \nonumber
	M\sbr{*} &= M_{xx}+M_{yy}+2\sqrt{\left|\mathbf{M}\right|} \mathrm{.}
\end{align}
As the dipole moments of the PSF are used to determine its centroid, it is important to test them as well, so we will additionally include $M_x$ and $M_y$ in our analysis.

From work by \citet{PauAmaVoi08} and \citet{Row10}, we know in particular that errors in the PSF size will lead to multiplicative biases in shear estimates, and errors in PSF shape will lead to additive biases. We can identify here $M\sbr{*}$ as related to the size of the PSF, and $M\sbr{+}$ and $M\sbr{\times}$ as related to the shape of the PSF, which implies errors in the former will cause multiplicative biases, and errors in the latter will cause additive biases. This is confirmed by our own calculations in \Appref{shear_bias_relation}.

Note that of these five terms, all except $M\sbr{*}$ are independent to changes to the background by a flat offset in the limit of zero pixel size, due to the cancellation of terms in the calculation. Even outside this limit, this will tend to be the case on average due to the uniform random positioning of stars relative to the centres of the nearest pixels. For the quadrupole terms (but not the dipole terms), this is also true for changes to the background by a linear gradient. This means that even if the observation has a non-zero background in the region surrounding a star used for comparison with the model, all quantities except for $M\sbr{*}$ will be relatively insensitive to this effect. However, as $M\sbr{*}$ is sensitive to this, a measured difference in it between the model and observed stars may be due either to imperfect flat-fielding or background subtraction, or to the model itself being imperfect. Worse, it is possible for the effects of imperfect flat-fielding or background subtraction to cancel out the effects of model imperfections, falsely reassuring us that the model is adequate. Therefore, either we must ensure that any imperfections in subtracting off the sky background are unbiased with respect to the locations of the stars we select, or else we must test a different quantity in its place.

For our work here, we choose the latter approach, developing an alternative quantity which we will use in the place of $M\sbr{*}$. The key requirement on this quantity is that it should in general be positively correlated with changes in $M\sbr{*}$ due to changes in the central flux distribution but be insensitive to the addition of a flat light distribution. As $M\sbr{*}$ can be seen as representative of the distribution's size, we look for an alternate size measurement which is background-independent.

To accomplish this goal, we generate a new size estimator $M\sbr{s}$ through the following algorithm:
\begin{itemize}
	\item Bin all pixels in the image by distance from the centroid, in bins of size 1 pixel.

	\item For each bin $i$ except the innermost, determine $W_i = \overline{I_{<i}} - \overline{I_i} $, where $\overline{I_i}$ is the mean flux of all pixels in this bin, and $\overline{I_{<i}}$ is the mean flux of all pixels interior to it. For the innermost bin, $W_0 = 0$. These $W$ are designed to be insensitive to the background level -- a flat background added to the image will have no effect on the calculated $W_i$.\footnote{This form was inspired by the formula for the signal from weak gravitational lensing around a mass distribution, the result of which is degenerate with the addition of a constant-value mass sheet to the field of view.} Due to the symmetry of this procedure, they will also be insensitive to any linear gradient in the background.
 
	\item Determine the size measure $m\sbr{s}' = \sum_i( d_i w(d_i) W_i ) / \sum_i( w(d_i) W_i )$, where $d_i$ is the distance from the centre of the image to the middle of each bin and $w(r)$ is the weight function used for other moments (which is limited to being circularly-symmetric for this method).

	\item Repeat the above for an image in which the flux of the central pixel is 1 and all other pixels are zero, to determine $m\sbr{s}^0$. Determine the final size estimate $M\sbr{s} = (m\sbr{s}' - m\sbr{s}^0)^2$. The subtraction here is performed to regularise $M\sbr{s}$ so that it converges to zero for objects of the minimum possible size, and it is squared so that it has equivalent units to $M\sbr{*}$.
\end{itemize}
This size estimator is analysed in detail in \Appref{Qsize}. Notably, we find that in the circumstances we intend to use it for, this estimator has the additional advantage that it is relatively less sensitive to noise than other commonly-used size estimators.

We now have the parameters $M\sbr{+}$, $M\sbr{\times}$, $M\sbr{s}$, $M_x$, and $M_y$, all of which we expect to be the same between an ideal model PSF and a selection of isolated stars, differing only due to noise. We will refer to this set of parameters collectively as $M_k$.

\subsubsection{Combining Weight Functions}
\Seclabel{comb_weight}

The parameters we developed in the previous section are sensitive to the choice of weight function used. If the weight function is strongly weighted toward the core of the PSF, these parameters will be relatively insensitive to imperfections outside the core, which might pose issues if galaxy ellipticities are ultimately measured with a weight function which is less weighted toward the core (with similar issues in the opposite scenario). Many shear-measurement algorithms use adaptive weight functions, which makes it impossible to know which weight function to use here. What we can do is to bracket the possible weight functions, using one weight function which is weighted to the core of the PSF, and another which is weighted to the wings. For our tests, we use the following weight functions:
\begin{align}
	\mathrm{Core:}\;w\sbr{c}(R) &= \begin{cases}
	           	 \exp\left( -R^2/ (2\sigma\sbr{c}^2) \right) & R \leq R\sbr{max}, \\
	           	 0 & R > R\sbr{max},
	           \end{cases} \\  \nonumber
	\mathrm{Wings:}\;w\sbr{w}(R) &= \begin{cases}
	           	 1 & R \leq R\sbr{max}, \\  \nonumber
	           	 0 & R > R\sbr{max},
	           \end{cases}
\end{align}
with $\sigma\sbr{c}^2 = 0.15$ square arcseconds and $R\sbr{max} = 0\farcs5$. Note that the number of pixels within an annulus scales linearly with $R$, so for the wings weight function, the total influence of all pixels at radius $R$ will also scale with $R$, up to the cut-off radius $R\sbr{max}$.

An ideal PSF model will have all five of our tested $M_k$ parameters being on average the same for the models as for the star images for both the core and wings weight functions. However, the fact that these weight functions sample overlapping regions means that they are not independent of each other, and so we must take into account the resulting covariances of each $M$ parameter for the two weight functions in our final statistic for the model. We can do this by defining and using the quantities ``$Q^{(+)}_k$'' and ``$Q^{(-)}_k$'', which are linear combinations of the $M_k$ parameters as measured with the two weight functions:
\begin{align}
	\Eqlabel{Q_definitions}
	Q^{(+)}_k = \frac{1}{2\sqrt{2}}\Bigg[&M^{\rm (c)}_k \left(1+\frac{\sigma(M^{\rm (w)}_k)}{\sigma(M^{\rm (c)}_k)}\right) \\ \nonumber
	+ &M^{\rm (w)}_k \left(1+\frac{\sigma(M^{\rm (c)}_k)}{\sigma(M^{\rm (w)}_k)}\right)\Bigg] \mathrm{,} \\ \nonumber
	Q^{(-)}_k = \frac{1}{2\sqrt{2}}\Bigg[&M^{\rm (c)}_k \left(1+\frac{\sigma(M^{\rm (w)}_k)}{\sigma(M^{\rm (c)}_k)}\right) \\ \nonumber
	- &M^{\rm (w)}_k \left(1+\frac{\sigma(M^{\rm (c)}_k)}{\sigma(M^{\rm (w)}_k)}\right)\Bigg] \mathrm{,} \\ \nonumber
\end{align}
where $M^{\rm (c)}_k$ is the $M_k$ parameter using the core weighting function, $M^{\rm (w)}_k$ is the $M_k$ parameter using the wings weighting function, and $\sigma(M^{\rm (c)}_k)$ and $\sigma(M^{\rm (w)}_k)$ are the standard deviations of these quantities. The covariance of these two quantities can be shown to be zero, and in the scenario where $\sigma(M^{\rm (c)}_k)=\sigma(M^{\rm (w)}_k)$, we will have the property $\left(Q^{(+)}_k\right)^2 + \left(Q^{(-)}_k\right)^2 = \left(M^{\rm (c)}_k\right)^2 + \left(M^{\rm (w)}_k\right)^2$, implying the $Q^{(\pm)}_k$ values will be of similar magnitude to the $M^{\rm (c/w)}_k$ values.

Using these linear combinations has the additional benefit that it allows us to address the fact that the PSF centres are determined as the positions which make the dipoles zero for a certain weight function. If we were using a single weight function, it would be impossible to detect any differences between the dipole moments of the model and observed point sources, as these differences would be removed in the centring step. However, when we are using two weight functions, only the dipoles using one of them will be set to zero, and we can get information from the dipoles which use the other weight function.

If the weight function is flat near the centre (as is true of both tophat and Gaussian weight functions), then changes in the dipole moments will be linearly correlated with changes in the centre to a first-order approximation. Thus, if we shift the centre so that, for instance, $M^{\rm (c)}_{x}$ and $M^{\rm (c)}_{y}$ are made equal to zero, $M^{\rm (w)}_{x}$ and $M^{\rm (w)}_{y}$ will be reduced by an amount proportional to the original $M^{\rm (c)}_{x}$ and $M^{\rm (c)}_{y}$. The final $M^{\rm (w)}_{x}$ and $M^{\rm (w)}_{y}$ values will then be analogous the differences between the values for each weight function. Looking at the linear combinations we presented above, we see that $Q^{(-)}_{x/y}$ also uses a difference between the moments from the two weight functions. We can therefore still use $Q^{(-)}_{x/y}$ for our analysis. This gives as a final set of eight values which provide a linearly independent basis: $Q^{(-)}_{x}$, $Q^{(-)}_{y}$, $Q^{(+)}_{+}$, $Q^{(-)}_{+}$, $Q^{(+)}_{\times}$, $Q^{(-)}_{\times}$, $Q^{(+)}_{s}$, and $Q^{(-)}_{s}$. In the following sections, we will use the shorthand ``$Q_k$'' to refer to this particular set of parameters.

\subsubsection{Quality of Fit Parameters}
\Seclabel{qof_parameters}

As many PSF models have one or more free parameters (the focus-secondary-mirror offset in the case of Tiny Tim) which must be fit for each image, it is useful to determine a single value which can serve as a quality-of-fit metric. The focus parameter can then be fit by minimising this value. A natural value to use for this purpose is the $\chisq$ value for the $Q_k$ parameters, but this would require us to calculate a theoretical estimate of the errors for each $Q_k$ value. This is not a trivial matter, as the determination of the centre of each star interacts with the calculation of the moments in a complicated manner. We therefore wish to use empirical estimates of the error in each parameter, calculating it from the scatter of the differences between the star and model values of that parameter for each image.

This, however, raises the issue that the resulting parameter from a $\chisq$-like calculation will not follow a standard $\chisq$ distribution, since the information in the scatter is reused both in the error calculation and in the $\chisq$ calculation. Additionally, a $\chisq$ value may not be optimal for fitting purposes: If a fit is imperfect, it indicates the statistical significance of a failure, rather than the practical significance of it. In our case, if we cannot fit perfectly, we would like to find the closest possibility, which may not be the same as that which minimises the statistical significance of the failure\footnote{Here, a notable difference between minimising the statistical and practical significances of fitting failures arises due to our decomposition into the $Q^{(+)}_k$ and $Q^{(-)}_k$ values. As $Q^{(i)}_k$ represents a difference between highly correlated values, the statistical error on it is much smaller than the sum, $Q^{(+)}_k$. As such, if we were to take a $\chisq$-like value, weighting the differences from zero by their errors, a similar practical difference (eg. $(0.1,0.1)$ versus $(0.1,-0.1)$) will correspond to a much larger statistical significance if it corresponds to a non-zero $Q^{(-)}_k$ value than if it corresponds to a non-zero $Q^{(+)}_k$ value.}. We can calculate a value more indicative of the practical significance through the following process. We start by calculating $Z^2_k$ values through
\begin{align}
	\Eqlabel{Z2_def}
	Z_x^{2(-)} &= \frac{(0.71 \px^{-2})^2}{N}\sum_{i=0}^{N}\left(Q_{x,{\rm star},i}^{(-)}-Q_{x,{\rm model},i}^{(-)}\right)^{4} \mathrm{,} \\ \nonumber
	Z_y^{2(-)} &= \frac{(0.71 \px^{-2})^2}{N}\sum_{i=0}^{N}\left(Q_{y,{\rm star},i}^{(-)}-Q_{y,{\rm model},i}^{(-)}\right)^{4} \mathrm{,} \\ \nonumber
	Z\sbr{+}^{2(\pm)} &= \frac{(0.14 \px^{-2})^2}{N}\sum_{i=0}^{N}\left(Q_{{\rm +},{\rm star},i}^{(\pm)}-Q_{{\rm +},{\rm model},i}^{(\pm)}\right)^2 \mathrm{,} \\ \nonumber
	Z\sbr{\times}^{2(\pm)} &= \frac{(0.14 \px^{-2})^2}{N}\sum_{i=0}^{N}\left(Q_{{\rm \times},{\rm star},i}^{(\pm)}-Q_{{\rm \times},{\rm model},i}^{(\pm)}\right)^2 \mathrm{,} \\ \nonumber
	Z\sbr{s}^{2(\pm)} &= \frac{(0.43 \px^{-2})^2}{N}\sum_{i=0}^{N}\left(Q_{{\rm s},{\rm star},i}^{(\pm)}-Q_{{\rm s},{\rm model},i}^{(\pm)}\right)^2 \mathrm{.} \\ \nonumber
\end{align}
and a final fitting statistic:
\begin{equation}
	\Chisq = \sum_k Z^2_k \mathrm{,}
	\Eqlabel{Chisq_def}
\end{equation}
summing over the set of eight $Z^2_k$ values. (Here we use ``$Z^2_k$'' as a shorthand for this set of eight parameters.) The formulae here are justified in \Appref{shear_bias_relation}, but in summary:
\begin{itemize}
	\item $Z_x^{2(-)}$ and $Z_y^{2(-)}$ approximate the worst-case scenario for the mean square of contributions to shear bias from centroiding issues
	\item $Z\sbr{+}^{2(\pm)}$ approximates the mean square contribution of PSF shape inaccuracies to the first additive component of shear bias ($c\sbr{1}$)
	\item $Z\sbr{\times}^{2(\pm)}$ approximates the mean square contribution of PSF shape inaccuracies to the second additive component of shear bias ($c\sbr{2}$)
	\item $Z\sbr{s}^{2(\pm)}$ approximates the mean square contribution of PSF size inaccuracies to the both multiplicative components of shear bias ($m\sbr{1}+m\sbr{2}$)
	\item $\Chisq$ approximates the mean square of total contribution to shear bias
\end{itemize}
We can minimise $\Chisq$ to fit the optimal PSF model, but due to the presence of noise in the images, if we wish to judge whether or not a given image passes a test of the PSF model, we will need a baseline for its expected value and variance in ideal circumstances. We accomplish this by applying our testing procedure to a set of control images, as detailed in \Secref{Control_Tests} below.

The choice of the value $\Chisq$ to minimise is admittedly arbitrary, and an argument could be made that it is better to weight the contributions to additive and multiplicative biases differently, given the different magnitudes of and requirements on them. However, it is first necessary to see if one or the other might be more problematic before choosing such weights, and so we use equal weights in our analysis here, and invite others to let our results inform how they might choose to weight these components.

It is also worth comparing to requirements that might be imposed on the PSF. For instance, HST images are planned to be used for validation of the processing pipeline for the Euclid mission \citep{LauAmiArd11}. This mission imposes the following shear bias requirements on its full pipeline:
\begin{align}
	m\sbr{1}\mathrm{,}\;m\sbr{2} &\leq 2 \times 10^{-3}\mathrm{,} \\ \nonumber
	c\sbr{1}\mathrm{,}\;c\sbr{2} &\leq 5 \times 10^{-5}\mathrm{.} \\ \nonumber
\end{align}
The requirements on HST images for validation of the Euclid pipeline are likely to be much less strict, but as these requirements have not yet been calculated, we will use the above values as example requirements for our work here. This will allow us to judge, for instance, whether the multiplicative or additive requirements on shear bias are more difficult for PSF models to meet.

These shear bias requirements will correspond to requirements on the various $Z^2_k$ values. However, in order to test against these requirements, we must first account for the fact that noise in observations translates to non-zero typical $Z^2_k$ values (which are in effect scaled mean square deviates) even for an ideal PSF model. If we assume that the $Z^2_k$ values we calculate for the control fields in the ideal scenario, which we will here label as $Z^2_{k,{\rm Ideal}}$, then we can consider the $Z^2_k$ values for observations to be the sum of this, a contribution from excess variance (which might, for instance, be caused by the model not fully capturing the spatial dependence of the PSF) and a contribution from a difference in the mean of the measured moments between the modelled and observed PSFs. That is,
\begin{equation}
	Z^2_k = Z^2_{k,{\rm Ideal}} + \mathit{Z}^2_{k,{\rm Excess Variance}} + \mathit{Z}^2_{k,{\rm Mean Deviate}}\mathrm{.}
\end{equation}
It is only the latter two terms which we want to test against requirements on the PSF, which we can do by imposing the requirements on the difference between the measured $Z^2_k$ values and those measured for the control images, $Z^2_{k,{\rm Ideal}}$.

Additionally, if we wish to ensure that we meet requirements to a given threshold of certainty (for instance, $95$ per cent), we must require that this difference is below the required value by at least $1.645\sigma\left(Z^2_{k}\right)$, where the factor $1.645$ is the z score which corresponds to a one-sided p-value of $0.95$, and $\sigma\left(Z^2_{k}\right)$ is the standard deviation of this $Z^2_k$ value, as measured from tests on control images. This gives us the requirements:
\begin{align}
	Z\sbr{+}^{2(\pm)} - Z\sbr{+,Ideal}^{2(\pm)} &\lesssim 2.5 \times 10^{-9} - 1.645\sigma\left(Z\sbr{+}^{2(\pm)}\right) \mathrm{,} \\ \nonumber
	Z\sbr{\times}^{2(\pm)} - Z\sbr{\times,Ideal}^{2(\pm)} &\lesssim 2.5 \times 10^{-9} - 1.645\sigma\left(Z\sbr{\times}^{2(\pm)}\right) \mathrm{,} \\ \nonumber
	Z\sbr{s}^{2(\pm)} - Z\sbr{s,Ideal}^{2(\pm)} &\lesssim 4 \times 10^{-6} - 1.645\sigma\left(Z\sbr{s}^{2(\pm)}\right) \mathrm{.} \\  \nonumber
\end{align}
Here we assume that centroiding is handled carefully so that even if issues are present with the $x$ and $y$ moments of the PSF, they do not impact shear estimation. Note that these requirements are approximate due to the assumptions we made in our calculations of the weighting factors for the $Z^2_k$, which assume a typical galaxy size of $1.5\times$ the size of the PSF, and the specifics of the validation procedure used might require more or less stringent requirements.

\subsection{Implementing Tests}
\Seclabel{implementing_tests}

We implement our tests on each image independently. For each image, we start by using the SExtractor utility \citep{BerArn96} to identify all candidate objects in the image, using the image's exposure time to set the proper zeropoint for it in the configuration. To form our star sample, we then impose the following cuts on the generated catalogue:
\begin{itemize}
	\item CLASS\_STAR $\geq 0.99$
	\item $22.4 \leq$ MAG\_AUTO $\leq 25.4$ (approximately signal-to-noise $20$ to $200$)
	\item FLUX\_AUTO/FLUXERR\_AUTO $\geq 50$
\end{itemize}
This ensures that all objects in the sample are likely stars and are detected with enough significance to be useful for our tests. The lower limit on the magnitude is used to prune stars which are possibly bright enough to show non-linear effects or saturate their central pixels. For each star in the sample, we then determine the distance to the nearest other object (star or otherwise) in the catalogue. We remove any stars from the sample for which the nearest neighbour is within $2R\sbr{max}$ ($1\arcsec$) of it or the edge of the image is within $R\sbr{max}$ of it to form a sample of isolated stars. It is still possible that some of these stars might neighbour a faint object which was not identified by SExtractor or that they might in fact be a binary system, so it will be necessary later to check the sample for outliers.

For each of the isolated stars in our sample, we first determine the ideal centre position $(x\sbr{c},y\sbr{c})$ for it as the position for which $M^{\rm c}_{x} = M^{\rm c}_{y} = 0$ (where the $c$ superscript represents calculation using the core weight function). To do this, we start with the centre provided by SExtractor and calculate $M^{\rm c}_{x}$ and $M^{\rm c}_{y}$. We then shift the centre positions by these values:
\begin{align}
	x_{{\rm c},{\rm new}} = x_{{\rm c},{\rm old}} + M^{\rm (c)}_{x}\mathrm{,} \\ \nonumber
	y_{{\rm c},{\rm new}} = y_{{\rm c},{\rm old}} + M^{\rm (c)}_{y}\mathrm{.} \\ \nonumber
\end{align}
This process is iterated until the convergence of both $x\sbr{c}$ and $y\sbr{c}$. Using this centre, we then calculate the moments and $Q_k$ values for the star, as detailed in \Secref{relevant_quantities} and \Secref{comb_weight}.

We now proceed to determine the best-fit focus value for each image, by fitting for the focus value which minimises $X^2$ (as defined in \Eqref{Chisq_def}) for the image. The calculation of $X^2$ for each image is detailed in \Secref{qof_parameters}, except for the determination of outliers. To do this, we first calculate all $M_k^{\rm (c)}$ and $M_k^{\rm (w)}$ values (i.e. $M_{+}$ etc. calculated with each of the core and wings weight functions) except for $M^{\rm (c)}_x$ and $M^{\rm (c)}_y$ (which will be zero due to the centre fitting) for each star, assuming a focus offset value of $-1$ micron, which is at the centre of the observed range of fit values, $-8$ to $+6$ microns. For each of these parameters, we then calculate the mean and standard deviation of the differences between the measured values for the stars and their corresponding models.

Since we expect that there will be some contamination of our sample due to objects misclassified as stars, binary star systems, and other blends, we apply Chauvenet's criterion to each parameter, marking as outliers for this parameter stars where:
\begin{equation}
	\frac{ \left| (M_{k,{\rm star},i}-M_{k,{\rm model},i}) - \overline{(M_{k,{\rm star}}-M_{k,{\rm model}})}\right| }{\sigma(M_{k,{\rm star}}-M_{k,{\rm model}})} > D\sbr{max}(N)\mathrm{,}
\end{equation}
where $D\sbr{max}(N)$ is chosen such that for $N$ realisations of a normal distribution, there is a less than $50$ per cent chance that any value will be this far or farther from the mean, and $N$ is the initial number of stars. This process is iterated, updating the standard deviation and mean (but not $N$) until the sample has converged. A star will be considered an outlier if any of its $Q^c$ or $Q^w$ parameters is marked as an outlier.\footnote{This is indeed likely to reject more stars than necessary, but aside from giving us a smaller sample size to work with, this is unlikely to affect our final statistics, as they are based off of empirical scatter calculations on the samples of non-outlier stars.}

We use the samples of non-outlier stars to calculate the $X^2$ value for each image, and fit for the focus that provides the minimum such value. Since we are only fitting a single parameter, we apply a simple brute-force procedure, sampling focus values between $-6$ and $6$ microns at intervals of $1$ micron. We take the best of these values, and then use a downhill simplex method to determine the best focus to within $0.1$ microns. This gives us the best-fit focus value, and we use the statistics this value provides for our analysis.

\section{Control Tests}
\Seclabel{Control_Tests}

It is important to check our testing procedure against control images for various reasons. It allows us to validate that it works and has no apparent bugs, it informs us of the ideal fitting statistic $X^2$ for various scenarios, and it allows us to perform convergence tests.

\subsection{Control Image Design}
\Seclabel{Control_Image_Design}

Our control images are generated to match the properties which are shared by all our test images:
\begin{itemize}
	\item Dimensions: $4096 \times 2048$ pixels
	\item Pixel scale: $0\farcs05 /\px$
	\item Gain: $2.0 e^{(-)}/\mathrm{ADU}$
	\item Instrumental Zeropoint\footnote{per \texttt{https://hst-docs.stsci.edu/display/ACSIHB/9.2+Determining+Count+Rates+from+Sensitivities}}: $26.50$
\end{itemize}
We also make the following choices to make the image representative of one of the exposures in our sample, using the values from one of the exposures of NGC-0104, labeled as \texttt{jb6v09shq} in the archive:
\begin{itemize}
	\item Exposure time: $1298\sec$
	\item Pixel noise: Approximated by Gaussian distribution with $\sigma = 51 e^{(-)}/\px$. This is a conservative overestimate of the noise, to account for the contributions of the wings of bright stars in addition to the image background and read noise.
	\item Chip: $1$
\end{itemize}

We generate a set of $100$ images, with focus offset values distributed uniformly between $-6.0$ and $4.0$, which covers the typical range of values expected in observations. For each image, we generate $1000$ mock stars, with positions drawn from a uniform random distribution on the image and magnitudes drawn from a uniform random distribution between $22.4$ and $25.4$, and with appropriate shot noise applied to the images. Tiny Tim is called to generate a PSF model for each star (using its actual position, rather than the nearest grid point as done elsewhere in our analysis), subsampled at a factor of $10\times$, which is the maximum factor allowed by Tiny Tim. Unlike in our testing procedure, where we rebin models through simple summation after shifting, here we use the GalSim toolkit \citep{RowJarMan15} to interpolate and integrate the models. This is much slower, but it allows for more precise determinations of the rebinned PSF models. We will later test the accuracy of the faster simple summation approach for various subsampling factors in \Secref{Control_Convergence}. By default, each star is generated using the K7V spectrum in Tiny Tim.

We generate a set of variant control images, each accounting for different effects which would be difficult or time-consuming to account for in our testing procedure, to estimate the impact these effects would have on our testing procedure. We limit our analysis to representative cases of different effects, using the following image variants, with random seeding used to maintain the positions of stars between variants:
\begin{itemize}
	\item \textbf{Base:} All default options, as described above.
	\item \textbf{Binaries:} For a randomly-selected $30$ per cent of stars, an additional mock star is added to the image at a random position within a circle of radius $1\px$ around the star's position. The additional star's magnitude is drawn from the same distribution as the other stars.
	\item \textbf{Wide Binaries:} As Binaries, except in all cases where a star is selected to be a binary, the additional mock stars are added to the image within circles of radius $2\px$ around the stars' positions instead of circles of radius $1\px$.
	\item \textbf{1D Guiding Error:} To simulate the effects of guiding error \citep[see][section 5.2.3]{LucDesGon18}, each model PSF is convolved with a tophat profile of length $0.2\px$ in the x-dimension and length $0$ in the y-dimension.
	\item \textbf{2D Guiding Error:} Differential velocity aberration causes both an elongation of images similar to that of guiding error and a scale change \citep{PirColErb01}. The latter is small enough to be negligible, but the former is of concern. We simulate this in the extreme case where this effect is orthogonal to the elongation caused by guiding error by convolving each PSF model with a tophat profile of length $0.2\px$ in both the x- and y-dimensions.
	\item \textbf{Galaxy Background:} To test the impact of possibly unresolved blends, we add to the image a background consisting of a randomly-generated galaxy field, with size and magnitude distributions designed to approximate what would be observed in the F606W filter. Galaxies with magnitude $\lesssim 28$ are rendered.
	\item \textbf{Varying Spectral Type:} Rather than using the K7V spectrum for all stars, stars are generated with spectral types drawn from a random distribution. This distribution is a simple model which allows all types but is strongly weighted toward redder types, with $P(i) \propto i^3$, where $i$ is the index of the spectral type as provided to Tiny Tim. For reference, the default K7V spectral type is index 15 out of 17 total.
	\item \textbf{Full:} This variant includes the combination of effects from the Binaries, 2D Guiding Error, Galaxy Background, and Varying Spectral Type variants.
\end{itemize}
Two additional possible complicating factors are jitter and imperfect correction for charge transfer inefficiency \cite[``CTI'', see][]{MasSchCor14}. Jitter results in a blurring of scale $\lesssim 7\times 10^{-3} \arcsec$ \citep{Cla09}, or $0.14 \px$, which is sub-dominant to the guiding error and differential velocity aberration effects and not worth testing separately. CTI manifests as a blurring of images in the readout dimension. Aside from the fact that CTI is a nonlinear transformation, this is similar to the convolution we use for our Guiding Error variants, and so it is not necessary to use separate variants for imperfect CTI correction.

\subsection{Convergence Tests}
\Seclabel{Control_Convergence}

\begin{figure*}
	\includegraphics[scale=0.64]{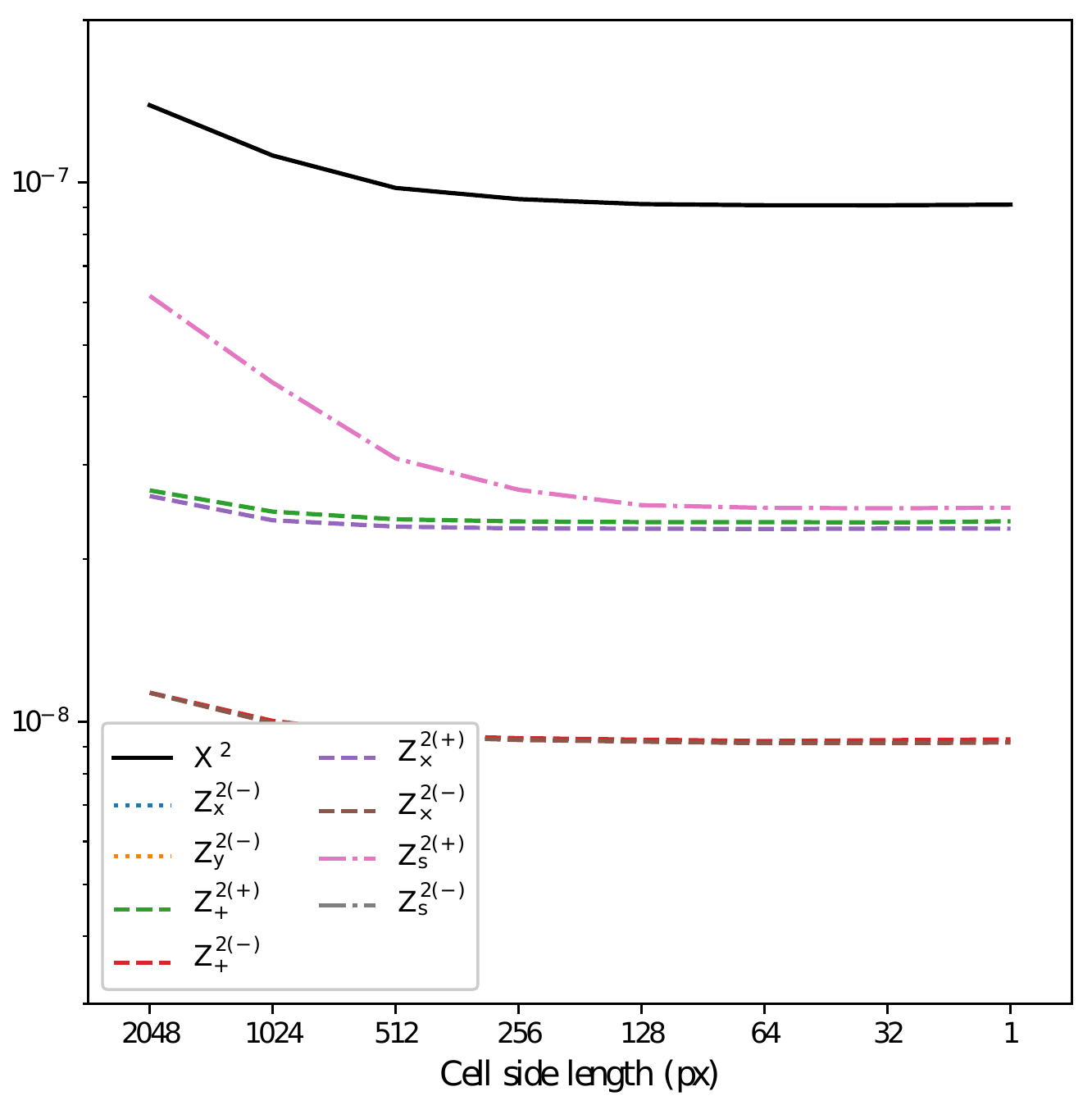}
	\includegraphics[scale=0.64]{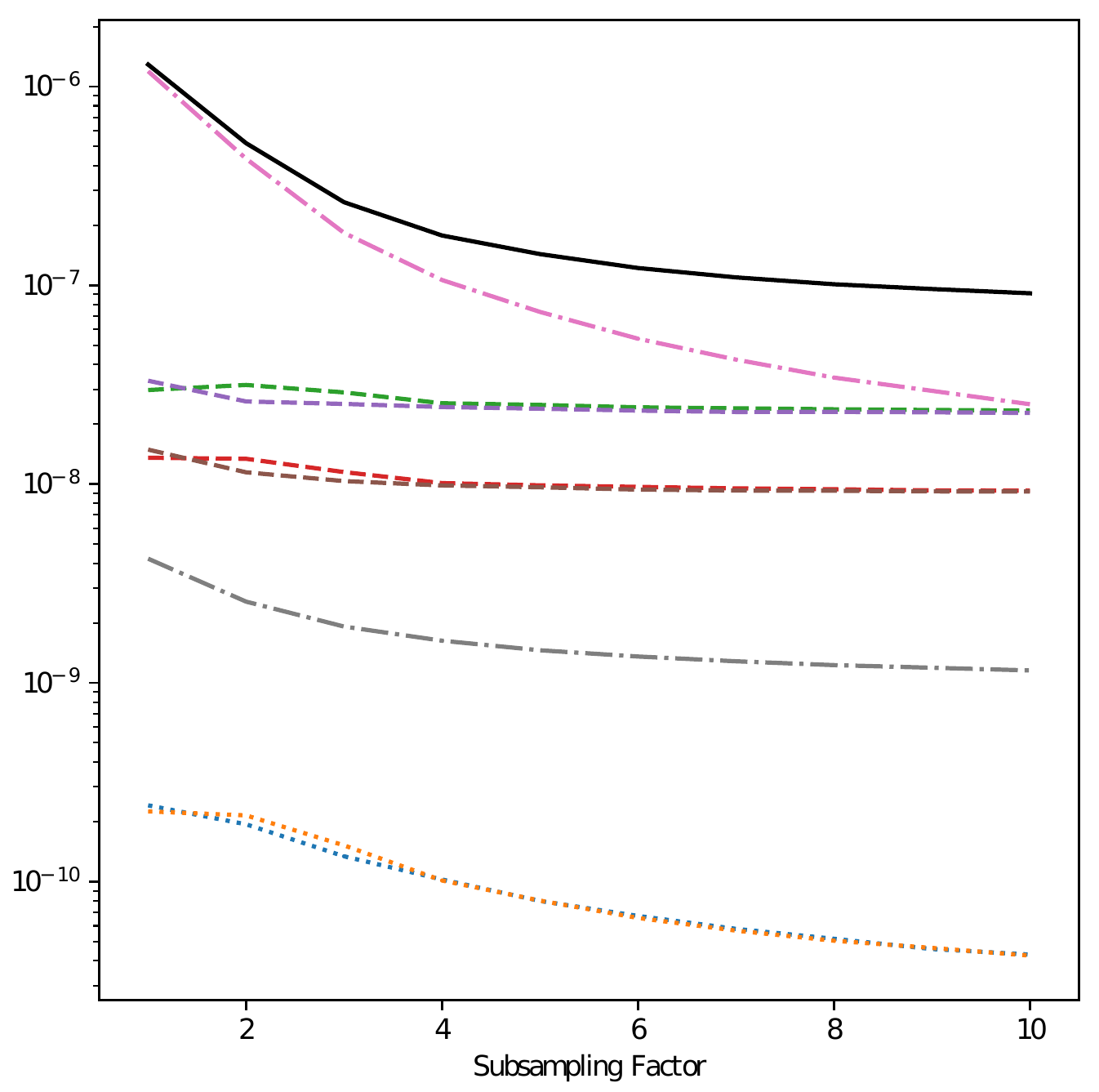}
	\caption{Convergence tests for the PSF gridding scheme (left) and subsampling factor (right). In each plot, the vertical axis shows the mean quality of fit statistics for a test of control fields using a given set up, where the solid black line shows $X^2$, the sum of the other statistics. In the left plot, the horizontal axis shows the side-length in pixels of each cell in the PSF gridding scheme, and in the right plot, it shows the subsampling factor used. Note that in the left plot, many lines aren't shown as scaling the plot to show them would result in making it difficult to judge the slope of the $\Chisq$ line, which is the most important result here.}
	\Figlabel{convergence_tests}
\end{figure*}

In \Figref{convergence_tests} we plot the results of our convergence tests. In the left plot, we test the gridding scheme for PSF models, where models are generated only for a fixed number of points at the centres of grid cells, and these models are used for all positions within their cells. This can greatly save time in testing if multiple stars reside in the same cell (even on different images), as fewer PSF models will need to be generated, but we must assess the impact of this approach to ensure that it does not significantly impact the results of our testing framework. We thus test the framework on our set of control images, using the known focus offset value for each and a set of cell sizes from $2048\px$ square (only two cells per image) to $1\px$ square. For all cases, we use the maximum possible subsampling factor of $10$ to ensure that no additional errors are introduced from a lack of convergence for that variable.

We see from the plot that if the gridding scheme is too coarse, $X^2$ can be increased by up to a factor of $\sim 1.5$. A cell size of $512\px$ square provides a reasonable balance of time saved versus accuracy, but for our purposes we decide to use a size of $128\px$ square. By this point, the accuracy has definitely converged, and it will likely still have converged if accuracy is improved (eg. by increased exposure time). This corresponds to $512$ cells per image, which means few stars on the same image will share a cell. It still allows for significant time savings if models are reused for different images, though.

In the right panel of \Figref{convergence_tests}, we test the subsampling factor used for the generated PSF models before they are shifted and rebinned to match the positions of stars. In this case, the effect on $X^2$ is more drastic, increasing it by up to an order of magnitude when no subsampling is used. Here, a subsampling factor of $4$ provides a good balance between time and accuracy. We choose to use a subsampling factor of $8$ for the same reasoning as before; it has converged by this point, and likely still will have if the accuracy is improved.

To ensure high accuracy for our tests, we choose a gridding scheme with cell sizes of $128\px$ square and a subsampling factor of $8$.

\subsection{Expected Fitting Statistics}
\Seclabel{Ex_X_Square}

\begin{table*}
	\caption{The fitting statistics resulting from applying our testing procedure to each control image variant. The details of the variants are listed in \Secref{Control_Image_Design}, and the fitting statistics are defined in \Secref{designing_tests}. The top table lists the fitting statistics resulting from a test in which the known focus offset for each control image is used, and the bottom table lists the statistics when the focus offset is fit for each image.}
\begin{center}
	\resizebox{\textwidth}{!}{\begin{tabular}{llllllllll}
		\multicolumn{10}{c}{\textbf{Known Focus Offset}} \\		Type & $X^2$ & $Z^{2(-)}_{x}$ & $Z^{2(-)}_{y}$ & $Z^{2(+)}_{+}$ & $Z^{2(-)}_{+}$ & $Z^{2(+)}_{\times}$ & $Z^{2(-)}_{\times}$ & $Z^{2(+)}_{\rm s}$ & $Z^{2(-)}_{\rm s}$ \\ \hline 
		Base & $1.0\times 10^{-7}$ & $5.1\times 10^{-11}$ & $5.1\times 10^{-11}$ & $2.4\times 10^{-8}$ & $9.5\times 10^{-9}$ & $2.3\times 10^{-8}$ & $9.2\times 10^{-9}$ & $3.5\times 10^{-8}$ & $1.2\times 10^{-9}$ \\
		Binaries & $1.6\times 10^{-7}$ & $3.7\times 10^{-11}$ & $3.9\times 10^{-11}$ & $2.0\times 10^{-8}$ & $8.6\times 10^{-9}$ & $1.9\times 10^{-8}$ & $8.6\times 10^{-9}$ & $9.9\times 10^{-8}$ & $1.5\times 10^{-9}$ \\
		Wide Binaries & $1.7\times 10^{-7}$ & $4.2\times 10^{-11}$ & $3.9\times 10^{-11}$ & $2.3\times 10^{-8}$ & $9.9\times 10^{-9}$ & $2.2\times 10^{-8}$ & $9.7\times 10^{-9}$ & $1.1\times 10^{-7}$ & $2.0\times 10^{-9}$ \\
		1D Guiding Error & $1.0\times 10^{-7}$ & $5.1\times 10^{-11}$ & $5.1\times 10^{-11}$ & $2.4\times 10^{-8}$ & $9.5\times 10^{-9}$ & $2.3\times 10^{-8}$ & $9.2\times 10^{-9}$ & $3.5\times 10^{-8}$ & $1.2\times 10^{-9}$ \\
		2D Guiding Error & $1.0\times 10^{-7}$ & $5.1\times 10^{-11}$ & $5.0\times 10^{-11}$ & $2.4\times 10^{-8}$ & $9.6\times 10^{-9}$ & $2.3\times 10^{-8}$ & $9.5\times 10^{-9}$ & $3.5\times 10^{-8}$ & $1.2\times 10^{-9}$ \\
		Galaxy Background & $1.0\times 10^{-7}$ & $5.7\times 10^{-11}$ & $5.8\times 10^{-11}$ & $2.4\times 10^{-8}$ & $9.6\times 10^{-9}$ & $2.3\times 10^{-8}$ & $9.2\times 10^{-9}$ & $3.4\times 10^{-8}$ & $1.2\times 10^{-9}$ \\
		Varying Spec. Type & $1.0\times 10^{-7}$ & $5.1\times 10^{-11}$ & $5.0\times 10^{-11}$ & $2.4\times 10^{-8}$ & $9.6\times 10^{-9}$ & $2.3\times 10^{-8}$ & $9.4\times 10^{-9}$ & $3.5\times 10^{-8}$ & $1.3\times 10^{-9}$ \\
		Full & $1.6\times 10^{-7}$ & $4.2\times 10^{-11}$ & $4.4\times 10^{-11}$ & $2.0\times 10^{-8}$ & $8.9\times 10^{-9}$ & $2.0\times 10^{-8}$ & $8.6\times 10^{-9}$ & $1.1\times 10^{-7}$ & $1.6\times 10^{-9}$ \\
	\end{tabular}}
\end{center}
\begin{center}
	\resizebox{\textwidth}{!}{\begin{tabular}{llllllllll}
		\multicolumn{10}{c}{\textbf{Fit Focus Offset}} \\		Type & $X^2$ & $Z^{2(-)}_{x}$ & $Z^{2(-)}_{y}$ & $Z^{2(+)}_{+}$ & $Z^{2(-)}_{+}$ & $Z^{2(+)}_{\times}$ & $Z^{2(-)}_{\times}$ & $Z^{2(+)}_{\rm s}$ & $Z^{2(-)}_{\rm s}$ \\ \hline 
		Base & $1.1\times 10^{-7}$ & $9.9\times 10^{-11}$ & $1.0\times 10^{-10}$ & $2.6\times 10^{-8}$ & $1.1\times 10^{-8}$ & $2.4\times 10^{-8}$ & $9.6\times 10^{-9}$ & $3.9\times 10^{-8}$ & $1.3\times 10^{-9}$ \\
		Binaries & $1.6\times 10^{-7}$ & $6.9\times 10^{-11}$ & $7.1\times 10^{-11}$ & $2.2\times 10^{-8}$ & $9.9\times 10^{-9}$ & $2.0\times 10^{-8}$ & $8.9\times 10^{-9}$ & $9.5\times 10^{-8}$ & $1.7\times 10^{-9}$ \\
		Wide Binaries & $1.8\times 10^{-7}$ & $7.0\times 10^{-11}$ & $7.0\times 10^{-11}$ & $2.4\times 10^{-8}$ & $1.1\times 10^{-8}$ & $2.3\times 10^{-8}$ & $1.0\times 10^{-8}$ & $1.1\times 10^{-7}$ & $2.4\times 10^{-9}$ \\
		1D Guiding Error & $1.1\times 10^{-7}$ & $1.0\times 10^{-10}$ & $1.1\times 10^{-10}$ & $2.5\times 10^{-8}$ & $1.1\times 10^{-8}$ & $2.4\times 10^{-8}$ & $9.5\times 10^{-9}$ & $3.9\times 10^{-8}$ & $1.3\times 10^{-9}$ \\
		2D Guiding Error & $1.1\times 10^{-7}$ & $1.0\times 10^{-10}$ & $1.0\times 10^{-10}$ & $2.5\times 10^{-8}$ & $1.1\times 10^{-8}$ & $2.4\times 10^{-8}$ & $9.8\times 10^{-9}$ & $3.9\times 10^{-8}$ & $1.3\times 10^{-9}$ \\
		Galaxy Background & $1.1\times 10^{-7}$ & $1.1\times 10^{-10}$ & $1.1\times 10^{-10}$ & $2.6\times 10^{-8}$ & $1.1\times 10^{-8}$ & $2.4\times 10^{-8}$ & $9.4\times 10^{-9}$ & $3.8\times 10^{-8}$ & $1.3\times 10^{-9}$ \\
		Varying Spec. Type & $1.1\times 10^{-7}$ & $9.3\times 10^{-11}$ & $1.0\times 10^{-10}$ & $2.5\times 10^{-8}$ & $1.1\times 10^{-8}$ & $2.4\times 10^{-8}$ & $9.6\times 10^{-9}$ & $3.9\times 10^{-8}$ & $1.3\times 10^{-9}$ \\
		Full & $1.6\times 10^{-7}$ & $7.7\times 10^{-11}$ & $8.9\times 10^{-11}$ & $2.3\times 10^{-8}$ & $1.1\times 10^{-8}$ & $2.1\times 10^{-8}$ & $9.1\times 10^{-9}$ & $9.6\times 10^{-8}$ & $1.9\times 10^{-9}$ \\
	\end{tabular}}
\end{center}
	\Tablabel{control_field_test_statistics}
\end{table*}

In order to determine the expected fitting statistic $X^2$ for an ideal scenario, we apply our testing framework to each control field, using the converged grid scheme and subsampling factor determined in \Secref{Control_Convergence}. We test both using the known focus offset value and fitting for the best value, and we plot the resulting $X^2$ and $Z^2_k$ values in \Tabref{control_field_test_statistics}.

Let us first compare the fitting statistics from using the known focus offset versus fitting it, to understand the impact of the fitting procedure on the results. In the case where we use the known focus offset, the size-related parameter $Z^{2(+)}\sbr{s}$ is the largest contributor to $\Chisq$ in all scenarios, but particularly when unresolved binary stars are included. When the focus offset is instead fit, the magnitude of $Z^{2(+)}\sbr{s}$ tends to decrease slightly, while other statistics slightly rise, resulting in only a slight decrease in the total $\Chisq$. This is likely due to the fact that modifying the focus offset has the most significant impact on the size of the PSF model, and since this is already the largest contributor to $\Chisq$ - even when the observed size is incorrect (due to noise and possibly other factors) - $\Chisq$ can be improved by fitting a different focus offset. We will keep this effect in mind in our analysis.

\begin{figure*}
	\includegraphics[scale=0.75]{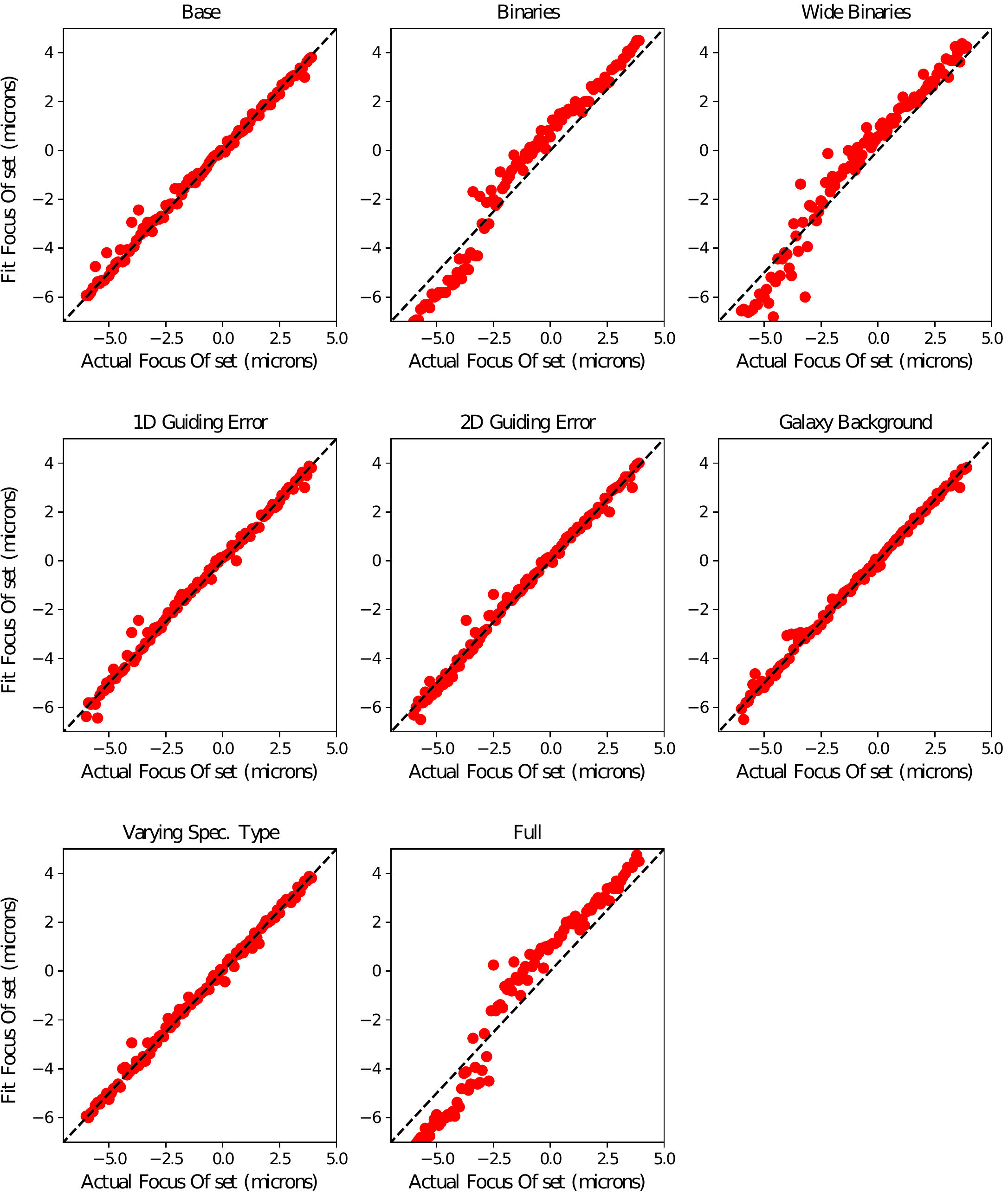}
	\caption{The fit focus offset plotted against the actual focus offset for each control image. Each panel shows a different control variant, including different complicating factors. The dashed line indicates an ideal fit, where all fitted focus offsets match the input values.}
	\Figlabel{focus_fitting_change}
\end{figure*}

In \Figref{focus_fitting_change}, we plot the fitted focus offset for each control image against the actual value for each control variant. We see that when unresolved binary stars are not included, the fitted value is very close to the actual value, but when binaries are included, the fitted values become more extreme, biased to either side of $\sim -3$. This is due to the minimum PSF size resulting from this value, and binaries biasing the fit to prefer larger model PSFs. Since larger PSFs result from focus offsets further from $\sim -3$, the fitted values are thus biased away from it, resulting in a bifurcation in the fitted values.

As the fit values are what we will obtain with observed images, we focus the remainder of our discussion on the values in that table. Looking first at the Base image variants, which include no extra complicating factors, we see that the quantities which impact additive shear bias, $Z^{2(\pm)}\sbr{+}$ and $Z^{2(\pm)}\sbr{-}$ are all significantly above the threshold to determine if the model PSFs meet the example requirements we have imposed, while those that impact multiplicative bias, $Z^{2(\pm)}\sbr{s}$, are well below the threshold. The standard deviations of these values (not shown in the table) are:
\begin{align}
	\sigma\left(Z^{2(+)}\sbr{+}\right) &= 3.2\times10^{-9}\mathrm{,}\\ \nonumber
	\sigma\left(Z^{2(-)}\sbr{+}\right) &= 1.6\times10^{-9}\mathrm{,}\\ \nonumber
	\sigma\left(Z^{2(+)}\sbr{\times}\right) &= 3.2\times10^{-9}\mathrm{,}\\ \nonumber
	\sigma\left(Z^{2(-)}\sbr{\times}\right) &= 1.6\times10^{-9}\mathrm{,}\\ \nonumber
	\sigma\left(Z^{2(+)}\sbr{s}\right) &= 7.2\times10^{-9}\mathrm{,}\\ \nonumber
	\sigma\left(Z^{2(-)}\sbr{s}\right) &= 2.2\times10^{-10}\mathrm{.}
\end{align}
Unfortunately, the standard deviations for the factors which contribute to additive bias on shear measurements are all sufficiently large they result in the threshold to confidently state that a PSF model meets this requirement being negative. This means that in practice we will not be able to confidently make this claim, although we may be able to confidently claim that a PSF model fails to meet this requirement. For the case of multiplicative bias, the standard deviation and ideal values of the contributing factors are sufficiently low that we can say that a model PSF confidently meets our multiplicative bias requirements if:
\begin{align}
	Z\sbr{s}^{2(+)} &\lesssim 4 \times 10^{-6} \mathrm{,\;and} \\  \nonumber
	Z\sbr{s}^{2(-)} &\lesssim 4 \times 10^{-6}\mathrm{.} \\ \nonumber
\end{align}

If it is in fact necessary to ensure the PSF model meets the example requirements for additive bias that we are using, testing only the brightest stars in an image will make this test more feasible mathematically, as it will reduce the impact of noise on the $Z^2_k$ values, but this will come at the expense of worse sampling of the PSF across the image. Taking deeper exposures can additionally help mitigate this problem. However, this applies only to the ideal scenario, which assumes that various other complicating factors are properly handled. It also assumes that there is no loss of data due to bright stars becoming oversaturated and useless, or if there is, it is counteracted by fainter stars becoming usable.

The remaining rows in \Tabref{control_field_test_statistics} show the representative impact of these effects, which we will now discuss.

The most significant complicating factor is the presence of unresolved binaries in the sample of stars. Although our testing procedure includes both a detection step (with cuts on SExtractor's \texttt{CLASS\_STAR} parameter and object size) and an outlier-rejection step, the presence of binary stars still increases the fitting statistic significantly, to $\sim 1.6-1.8 \times 10^{-7}$. This error cannot be reduced through longer exposures, but instead will require binary stars to be identified and excluded from the sample, for instance through analyses of their spectra as measured by supplementary observations. Interestingly, there is only a small difference in the fitting statistic when the width of binaries is increased, which is likely due to wider binaries being more likely to be identified in the outlier-rejection process and excluded from the sample.

All other factors tested (the presence of a background of resolved and unresolved galaxies, guiding error, and varying spectral type) have no significant impact on the fitting statistic.

For our analysis in this paper, we take the approach of ignoring these effects in our testing procedure and comparing the resulting fitting statistics to those for the Full variant.

\section{Results and Analysis}
\Seclabel{Results}

Having confirmed that our testing procedure works on control images and obtained data on the expected fitting statistics for an ideal PSF model in ideal and more realistic situations, we now move on to testing Tiny Tim PSFs on the HST observations we introduced in \Secref{Data}. In \Secref{Init_Testing_Results}, we present the results of testing Tiny Tim with its default configurations, and we test fitting more advanced configurations in \Secref{Further_TT_Testing}. We compare our results against \citet{diMakLal08} and \citet{NieLal10}'s model for HST's focus offset in \Secref{Model_Comparison}, and discuss our results in \Secref{Discussion}.

\subsection{Testing Results}
\Seclabel{Init_Testing_Results}

We can judge the quality of model PSFs by looking at the $\Chisq$ and $Z^2_k$ statistics of our tests, as defined in \Secref{qof_parameters}. For an ideal model in ideal circumstances, we would expect these values to be clustered around those found for the Base control images, as seen in the first row of \Tabref{control_field_test_statistics}. As multiple factors which we have not accounted for can complicate the analysis, even an ideal model will likely have larger quality-of-fit statistics. The final row of \Tabref{control_field_test_statistics} presents a reasonable estimate of the upper bound, using representative values for various possible complicating factors, and so comparisons against these values will present a more conservative test of the PSF models.

\begin{figure*}
	\includegraphics[scale=0.5]{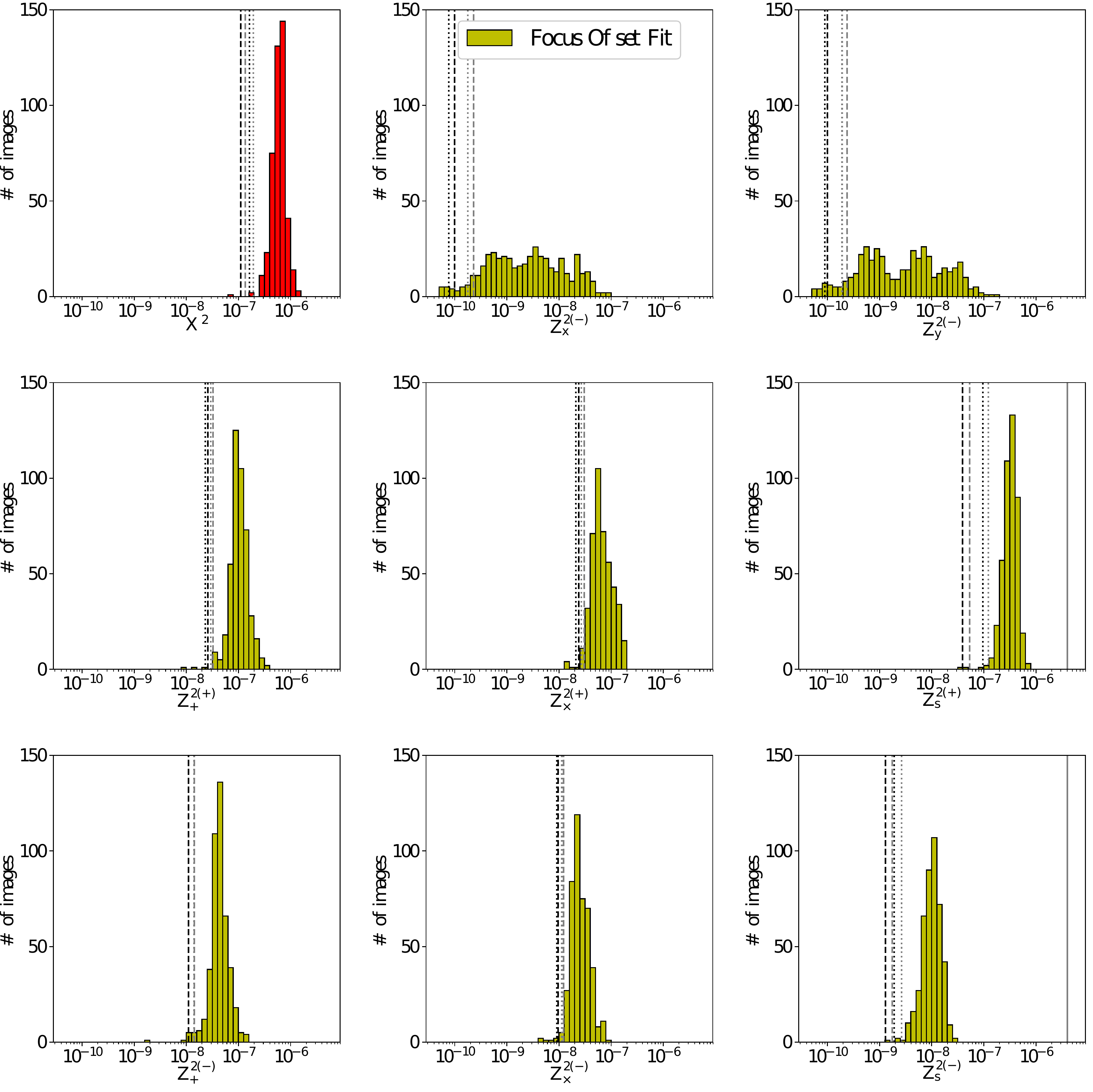}
	\caption[]{Distributions of the quality of fit parameter $X^2$ across all tested images, as the contributions of each tested moment, presented as $Z^2_k$ (see \Eqref{Z2_def}). The expected values for an ideal fit for each parameter are shown with the dashed black lines, and the expected values allowing for complicating factors are shown with the dotted black lines. The grey lines show these values plus twice their respective standard deviations among all control fields. The solid grey line, for the $Z^{2(+/-)}\sbr{s}$ panels, indicates the threshold for these values such that the PSF model will meet our example requirements on multiplicative bias $m < 2\times10^{-3}$. Our example requirements on additive bias are failed if the $Z^{2(+/-)}\sbr{+}$ or $Z^{2(+/-)}\sbr{\times}$ values exceed the control values.}
	\Figlabel{X2_inc_size_hists}
\end{figure*}

We plot histograms of our fitting statistics for all tested images after fitting the best focus offset for each in \Figref{X2_inc_size_hists}, and compare them to the expected values from the Base and Full control images, shown as the dashed and dotted black lines respectively. The corresponding grey lines show these values plus twice the standard deviation among all control fields, and so represents the approximate maximum values we would expect to see if the PSF models are ideal. If we look just at the total statistic, $X^2$, the values for the tested images are clustered significantly above the expected value from the Full control, which implies either an issue with the model or that our control underestimates the impact of complicating factors by nearly an order of magnitude.

This is similarly the case when we look at the $Z^2_k$ values which together compose $X^2$. Of these values, all are on average significantly above the expectations for the Full control. In the worst case, for $Z^{2(-)}_y$, the resulting statistics are on average nearly $1.5$ orders of magnitude higher than the expectation from the control images.

As the $Z^{2(+/-)}\sbr{+}$ and $Z^{2(+/-)}\sbr{\times}$ values generally greatly exceed those values for the control images, we can state that our example additive bias requirement ($c < 5 \times 10^{-5}$) will not be met by this PSF model. The multiplicative bias requirement will be met however, as although the $Z^{2(+/-)}\sbr{s}$ values generally exceed the control values, it is not by nearly enough to violate our requirement of $m < 2\times10^{-3}$.

\begin{figure*}
	\includegraphics[scale=0.59]{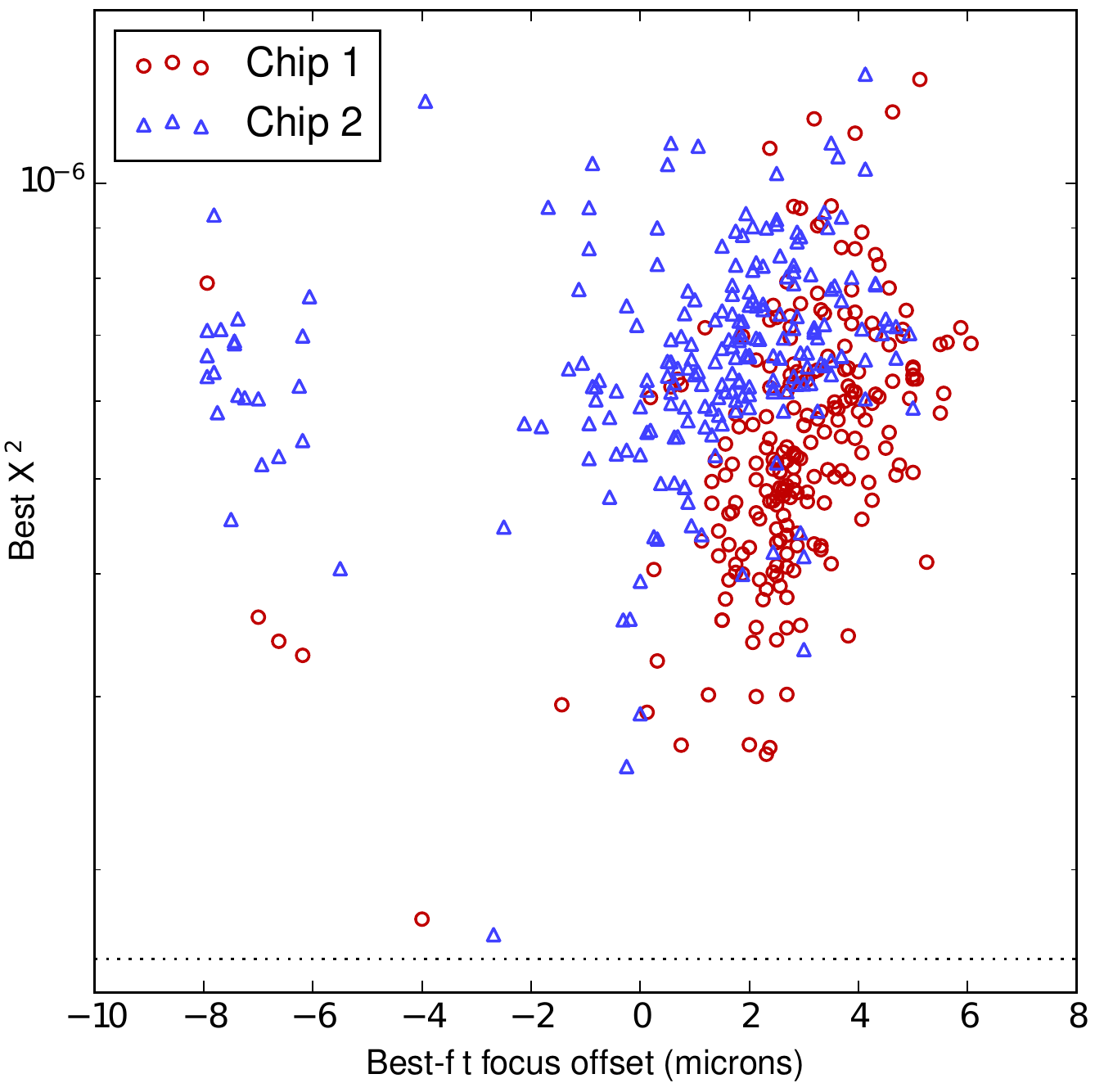}
	\includegraphics[scale=0.59]{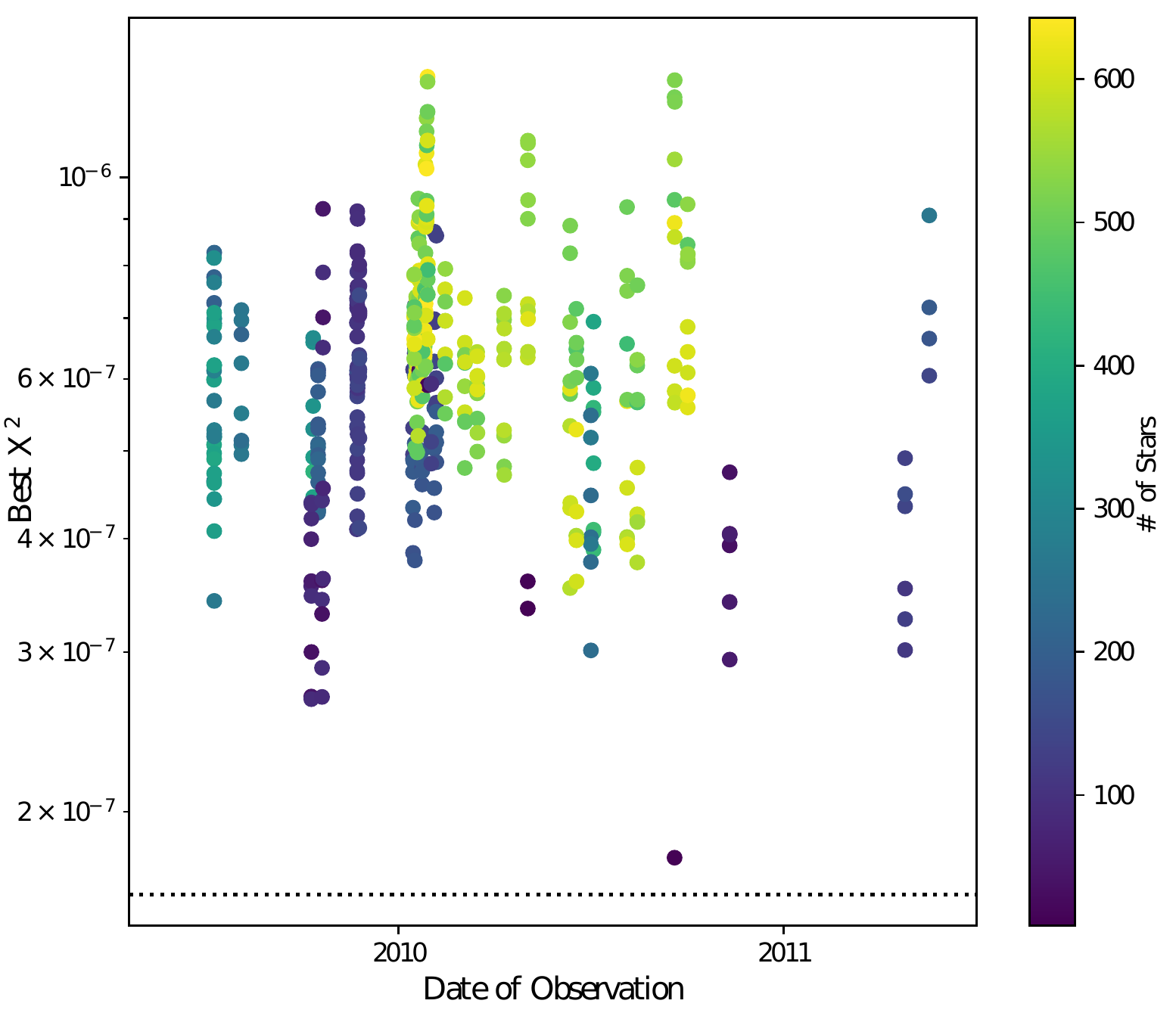}
	\caption{Left: The quality of fit parameter $\Chisq$ for each image plotted against the best-fit focus value, with points coloured according to which of the two detector chips the image was taken with. Right: The quality of fit parameter $\chisq$ for each image plotted against the date of observation, with points coloured according to the number of non-outlier isolated stars in the image.}
	\Figlabel{X2_v_other_inc_size}
\end{figure*}

We can gain further insight into the issue here by looking at the relationships between the quality of fit parameter $\Chisq$ and other parameters. We show this in \Figref{X2_v_other_inc_size}, where we plot the relationships between $\Chisq$ and the best-fit focus value, the chip with which each image was observed, the observation date, and the number of stars in each image. There is no apparent relationship between $\Chisq$ and the observation date, suggesting that there is no time-dependent effect in this period affecting the PSF (at least to our level of sensitivity). Similarly, if our exclusion of objects with nearby neighbours were not sufficient to deal with crowding, we would expect to see a trend between the number of stars in each image and $\Chisq$, and we see none. This suggests that our exclusion is indeed sufficient, at least for detected objects (and for undetected objects, the control tests outlined in \Secref{Ex_X_Square} show that this effect is minimal).

However, we do see notable oddities when we compare $\Chisq$ to the best-fit focus offset value and the chip with which each image was observed. We see that the focus offset values are clustered between $0$ and $6$ microns, where we expect from the HST's breathing that they will span the range of $-6$ to $6$ microns. The chip with which each image was observed also seems to play a role - images observed with chip 2 systematically have lower best-fit focus offset values by roughly a micron, and larger $\Chisq$.

One might consider that this is in part due to the fact that the fitting algorithm prioritises fitting the size of the PSF model, as it is the largest contributor to $\Chisq$. Since we cannot efficiently remove all unresolved binary stars from the sample, these bias the model to a focus offset value which provides a larger size. It seems that the fitting algorithm generally finds the best solution to this at larger focus offset values. However, this explanation is not fully consistent with our tests on control fields, as although the fitting does prefer more extreme focus offset values, as seen in \Figref{focus_fitting_change}, it is split between high and low values, with the mean bias actually being to a too-low focus offset. One potential explanation to this is that the symmetry observed in the control fields only holds for small deviations between the model and observed PSFs, and large deviations result in this symmetry breaking and appearing as we observe here.

Further analysis of the data shows that in the control fields, the model PSFs fail to match the sizes of the observed PSFs, but differences between the model and observed sizes are tightly clustered, while in the actual fields, the model PSFs match the sizes of the observed PSFs on average, but there is very large scatter in this relationship. This suggests that a possible explanation for this discrepancy might be that there is more spatial variation in the PSFs than is accounted for in the model. Inspection of the data shows that while there is a statistically significant correlation between position and size difference for many images, the nature of this correlation is not consistent between images. This could be due to temperature variations and gradients distorting the image plane in ways that are not accounted for by the Tiny Tim model.

The difference between the two chips might be considered to be due to the fact that there is a vertical offset between them. However, Tiny Tim already does implement a correction for this effect. Even if it didn't, a focus offset difference between the two chips of just one micron, as we tend to see here, would correspond to a height difference of hundreds of microns in the chips \citep{CoxNie11}, which is implausibly large. This effect was also noticed by \citet{CoxNie11}, who attributed it to most likely being due to differences in spherical aberration and charge diffusion between the two chips. As we will discuss further in \Secref{Further_TT_Testing}, we find this to be an insufficient explanation.

It seems that while the impact of unresolved binaries or some similar effect is responsible for some of the difference between our expected and measured $\Chisq$, this cannot explain all of it. Many of the fitting statistics are significantly worse in the real images than in our control tests, and this cannot be explained by any of the possible complicating factors we tested. The most likely conclusion at this point is that the issue lies with the model PSFs. In the following section, we will test options for improving the models.

\subsection{Advanced Configuration of Tiny Tim}
\Seclabel{Further_TT_Testing}

As we found in the previous section, the Tiny Tim PSF models we tested seem to fail to adequately characterise the observed PSFs, even when accounting for complicating factors that might result in apparently poor fits. However, the models we tested were limited to the default configuration for Tiny Tim. It is also possible to modify various parameters of the model, notably the Zernike polynomial coefficients which characterise the optics, but these options are not normally presented to the user. Given the discrepancies we have found between the models and reality, it is reasonable to consider that the default values for these parameters might be incorrect, at least for the observations we tested.

To test this hypothesis, we repeat our testing procedure, this time fitting each image not only for the focus offset (which in fact corresponds to the fourth Zernike polynomial's coefficient), but for the coefficients of all polynomials $Z_2$ through $Z_{21}$. Only the focus offset is expected to vary from image to image, so ideally the other fitted parameters will be consistent across all images, or else vary only with the date of observation (which would imply some factor changed the optics over time).

In addition to these, the possibility was also raised in private correspondence with Tiny Tim's author, John Krist, that the charge diffusion kernel determined by Tiny Tim is only an approximation and might not be accurate enough. As the possible variations to the kernel are infinite, we will test only one modification which will roughly characterise whether too much or too little intensity is diffused. We introduce a parameter $c$, and the eight non-central pixels of the kernel are then multiplied by $c$. The central pixel is then rescaled so that the kernel sums to $1$. A value of $c>1$ then corresponds to more diffusion, and a value $c<1$ corresponds to less diffusion. If there is an indication that the default value $c=1$ is not appropriate, further investigation can be undertaken to determine a better model for charge diffusion.

Finally, as we discussed in \Secref{Ex_X_Square}, it is possible to include the effects of guiding error and similar effects in the PSF model. This can be roughly approximated by convolving the PSF with a rectangular tophat profile, which will have free parameters representing its height, width, and orientation. However, this has the drawback that including this effect will necessitate a much greater amount of computer time: In addition to adding three more parameters which must be fit, this convolution would require the use of an interpolation and integration approach to rebinning PSF models, rather than simple summation. As this can result in up to double the total time required, and since the effect of guiding error is dwarfed by the effect of unresolved binary stars in the sample, we choose not to include this effect in our analysis.

We perform our fitting procedure using a two-stage approach. First, the focus offset is fit using a brute-force followed by steepest-descent algorithm while all other parameters are held constant. Secondly, a steepest-descent algorithm is used to fit all parameters. This is done to help avoid the fitting algorithm getting trapped in a local minimum. Since the focus offset is the most significant factor, fitting it first ensures that this approach will provide results at least as good as the focus-only approach, and will not get trapped in a local minimum far from the global minimum.

\begin{figure*}
	\includegraphics[scale=0.5]{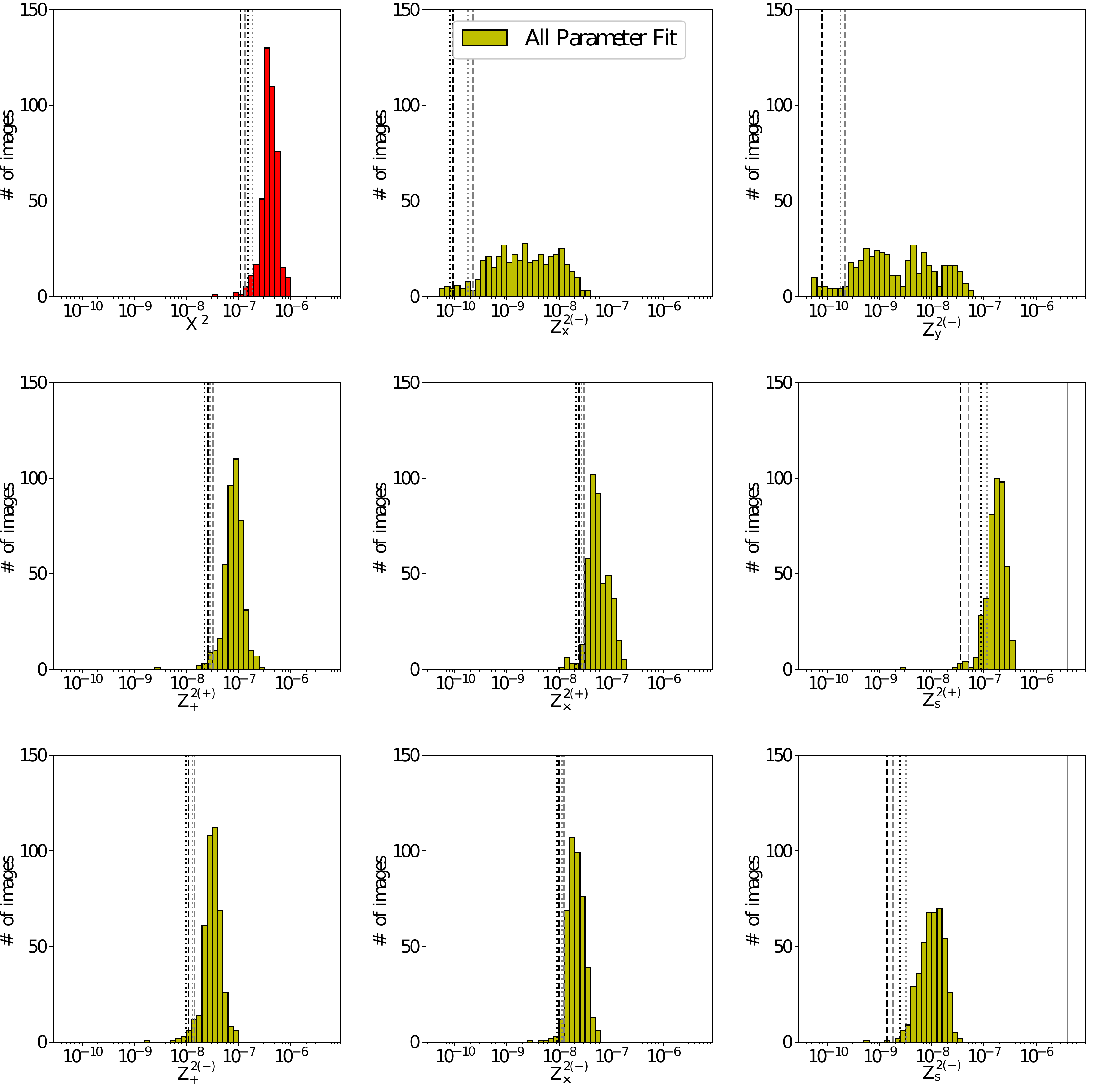}
	\caption[]{As \Figref{X2_inc_size_hists}, except all optical parameters were fit for each image.}
	\Figlabel{X2_all_params_hists}
\end{figure*}

In \Figref{X2_all_params_hists}, we plot the resulting fitting statistics when all optical parameters are fit for each image. This shows a small but noticeable improvement over the case where only the focus offset is fitted, and the overall statistic, $X^2$, lies closer to the mean value found when testing on representative control images. The fact that we still do not reach the expected quality-of-fit from the control tests might be due to our control images not properly accounting for some physical effect which is present in reality, or it might be due to a flaw in Tiny Tim's modelling.

The comparison to our example requirements remains the same even with this fit; the PSF model meets our requirements on multiplicative bias, but fails to meet our requirements on additive bias, although it is somewhat closer to meeting these requirements when all optical parameters are fit to each image. This is well beyond the accuracy limit to which Tiny Tim was originally tested, though, and is unlikely to be large enough to have any practical impact on shear measurements using HST data. It is only an issue here since we are testing the PSF model against the requirements for the Euclid mission, which are much stricter given the much greater volume of data involved.

\begin{figure*}
	\includegraphics[scale=0.59]{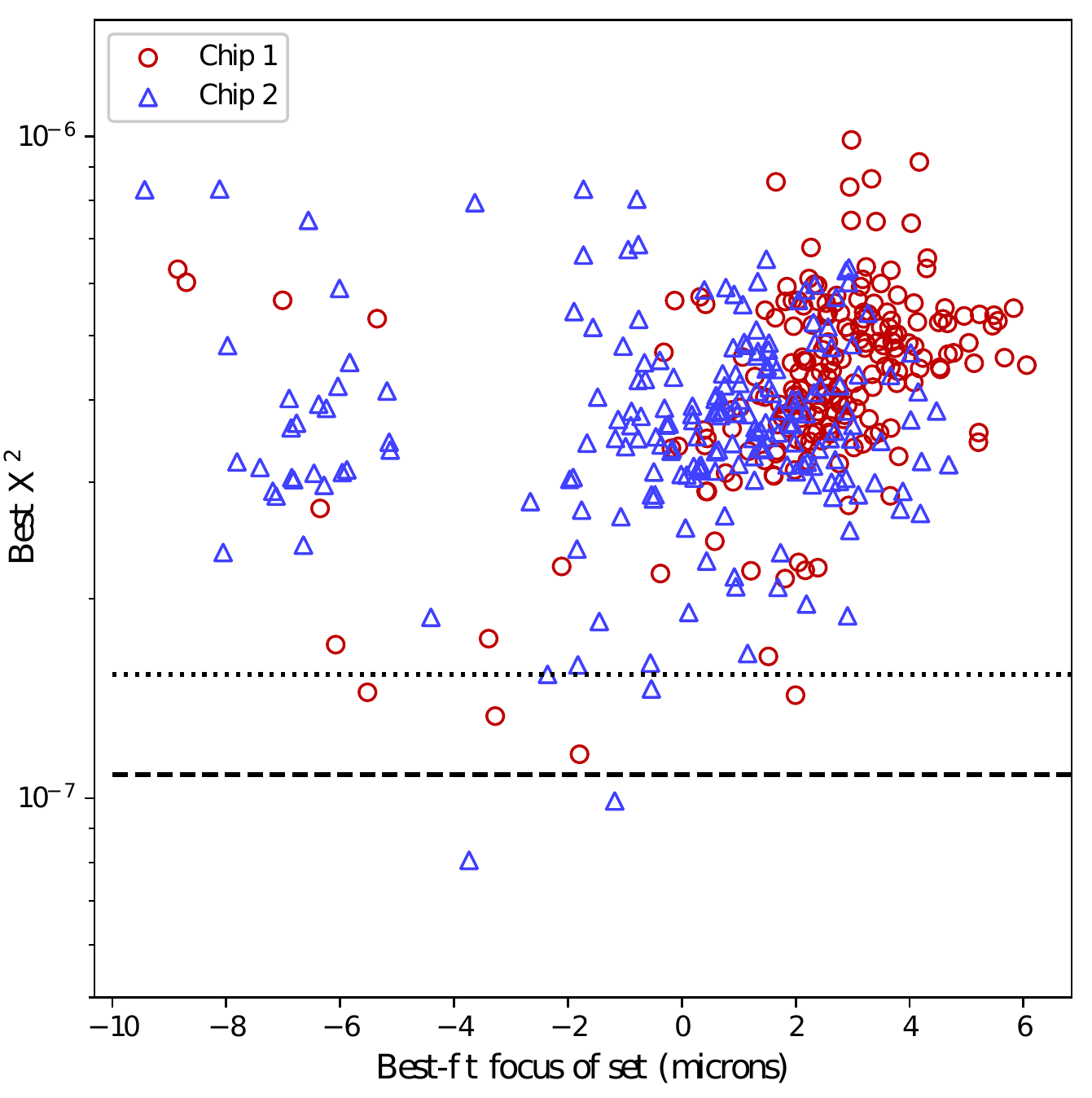}
	\includegraphics[scale=0.59]{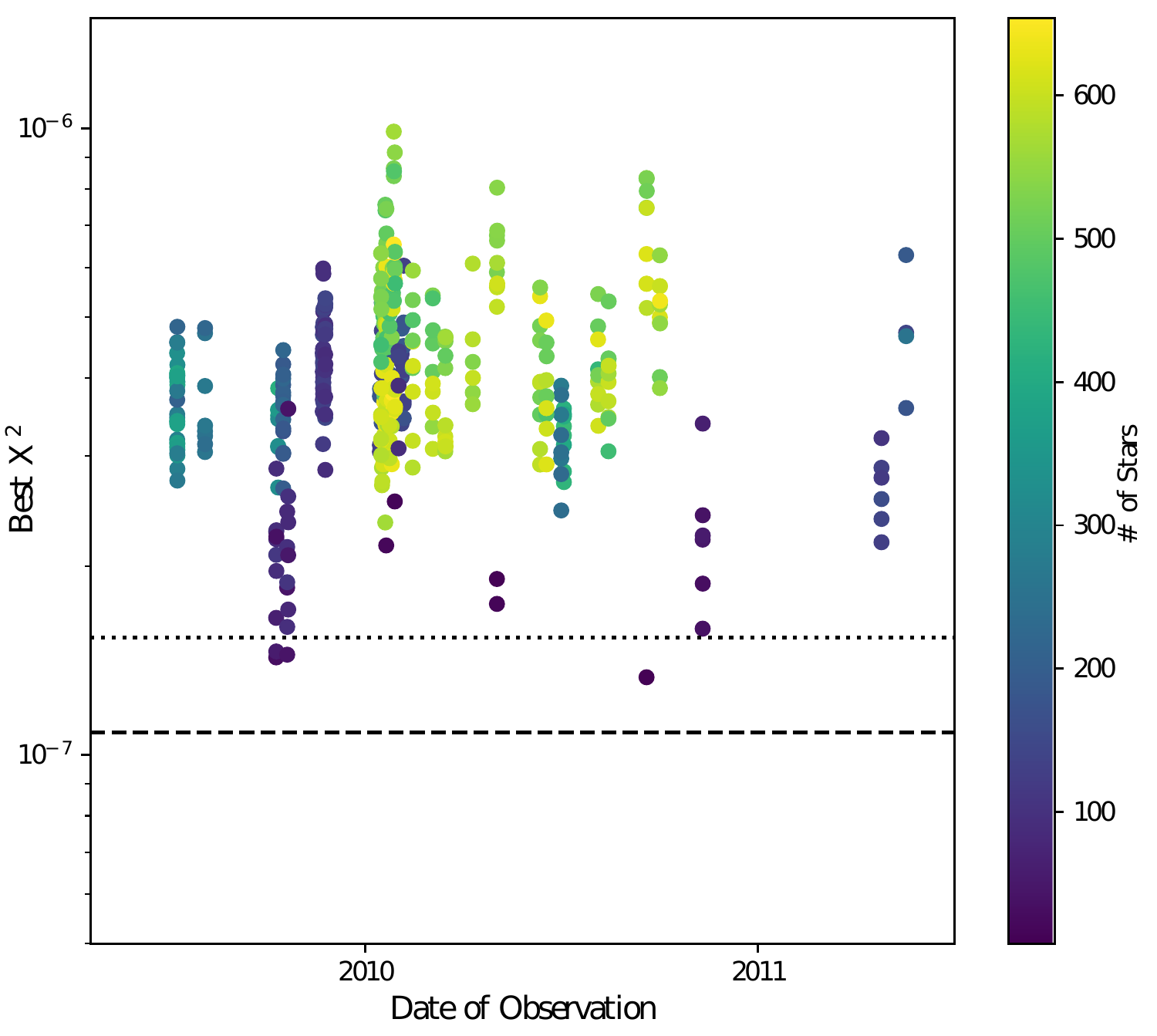}
	\caption{As \Figref{X2_v_other_inc_size}, except all optical parameters were fit for each image.}
	\Figlabel{X2_v_other_all_params}
\end{figure*}

The left panel of \Figref{X2_v_other_all_params} illustrates the relationship between the quality of fit and the fitted focus offset value now that all optical parameters are being fit. We no longer see a significant difference between the $\Chisq$ for the two chips, but a difference in the best-fit focus offset remains. This suggests that the issues we saw when only the focus offset was fit were possibly due to the use of other improper optical parameters being used. From the right panel, which shows the relationship between the quality of fit, the date of observation, and the number of stars per image, we see no significant change from before aside from the previously-noted fact that the quality of fit has generally improved.

\begin{figure*}
	\includegraphics[scale=0.68]{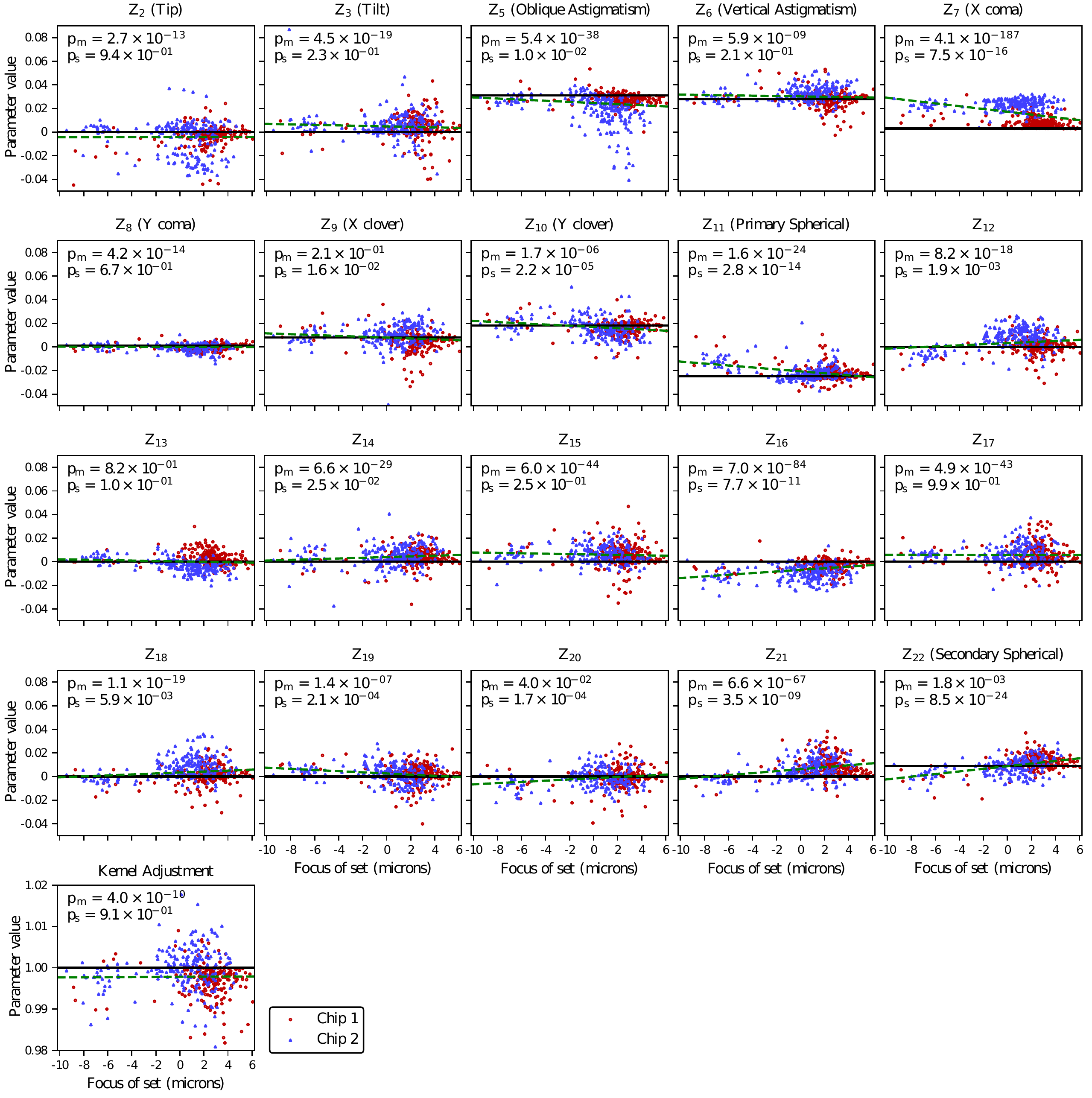}
	\caption{The best-fit optical parameters over all tested images, plus the adjustment $c$ to the charge diffusion kernel, plotted against the fitted focus offset value. The solid black line indicates the default value used for each by Tiny Tim, and the dashed green line is the best linear fit to the data. The blue triangular points correspond to data from Chip 1, and red circular points to data from Chip 2. The p-values stated on each panel are the two-tailed Gaussian probabilities that the mean fitted value is consistent with the default value and that the slope is consistent with zero. When the mean is separated from the default value by at least one standard deviation, or the slope is separated from zero by at least one standard deviation, the respective p-value is emboldened.}
	\Figlabel{bestfit_param_v_focus}
\end{figure*}

In \Figref{bestfit_param_v_focus}, we plot the best-fit values for each of these fitted parameters against the fitted focus offset, with Tiny Tim's default value for each shown for comparison with a solid black line. From this plot, we see that certain parameters are consistently fit to different values from the defaults used by Tiny Tim, and in some cases there is also a correlation between these fitted values and the fitted focus offset. Of particular note is the X coma parameter, which differs very significantly from its default value: Its mean fitted value is $0.0159$, compared to the default value of $0.003$.

It is possible that some or all of the cases where we see a correlation between these values and the focus offset might be due simply to a degeneracy in the models rather than a correlation between the actual values, but an actual correlation is possible. This could occur if the heating and cooling pattern of HST has more complicated effects than simply changing the focus offset. This could be caused by, for instance, certain parts of the telescope shading other parts, resulting in deformation due to different portions of it being heated by different amounts. This could result in the optical parameters for the PSF model being altered at the same time that the focus offset is altered, appearing as a linear correlation between them.

If our hypothesis is correct that the default optical parameters used by Tiny Tim are incorrect, then if we fit only the focus offset and use either the mean fitted values for the other parameters or the predictions from a linear regression of them against the focus offset, we should see significant improvement in the fitting statistics as compared to using the default parameters. If we do not see this, then it would imply that the improvement in fitting statistics here is likely due to fitting to noise, degeneracies between the focus offset and other parameters, and complicating factors such as unresolved binary stars.

\begin{figure*}
	\includegraphics[scale=0.5]{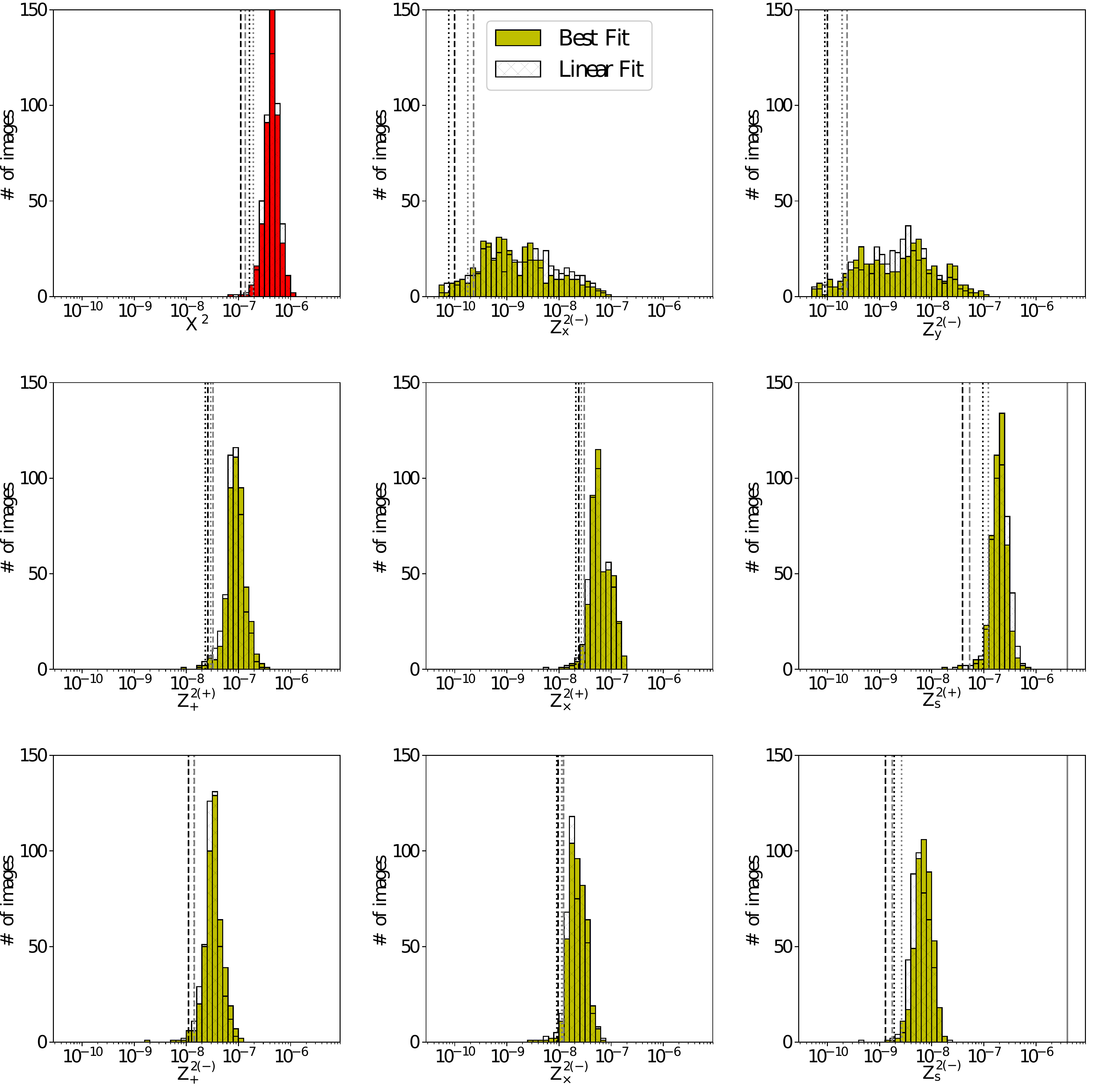}
	\caption[]{As \Figref{X2_inc_size_hists}, except showing cases where the mean best-fit optical parameters (aside from the focus offset) are used for the model PSFs (solid bars) and where a linear fit between the optical parameters and the focus offset is used (transparent bars). The focus offset is still fit for each image.}
	\Figlabel{X2_bestfit_params_hists}
\end{figure*}

In \Figref{X2_bestfit_params_hists}, we show the resulting fitting statistics from fitting the focus offset with either the mean best-fit optical parameters (solid bars) or a best-fit linear relationship between these parameters and the focus offset (transparent bars). Comparing this plot to \Figref{X2_inc_size_hists}, we see that either approach provides a notable improvement over using the default parameters, but not quite as much as fitting all optical parameters to each image. This also fails to bring the $\Chisq$ statistic below the expected value predicted from our representative control tests. This implies that the discrepancy between the models and reality cannot be explained entirely by the optical parameters having shifted since they were originally fit.

As using a linear relationship between the fitted focus offset and other parameters showed similarly little benefit, the discrepancy also cannot be adequately explained by these parameters varying due to the heating and cooling cycle of the instrument, like the focus offset does, unless they vary with a different periodicity or phase. We tested this hypothesis through Fourier analysis (see \Appref{params_FT}), but this failed to uncover any evidence of any parameters varying with a different periodicity or phase from the focus.

\subsection{Comparison to Focus Model}
\Seclabel{Model_Comparison}

\begin{figure}
	\includegraphics[scale=0.63]{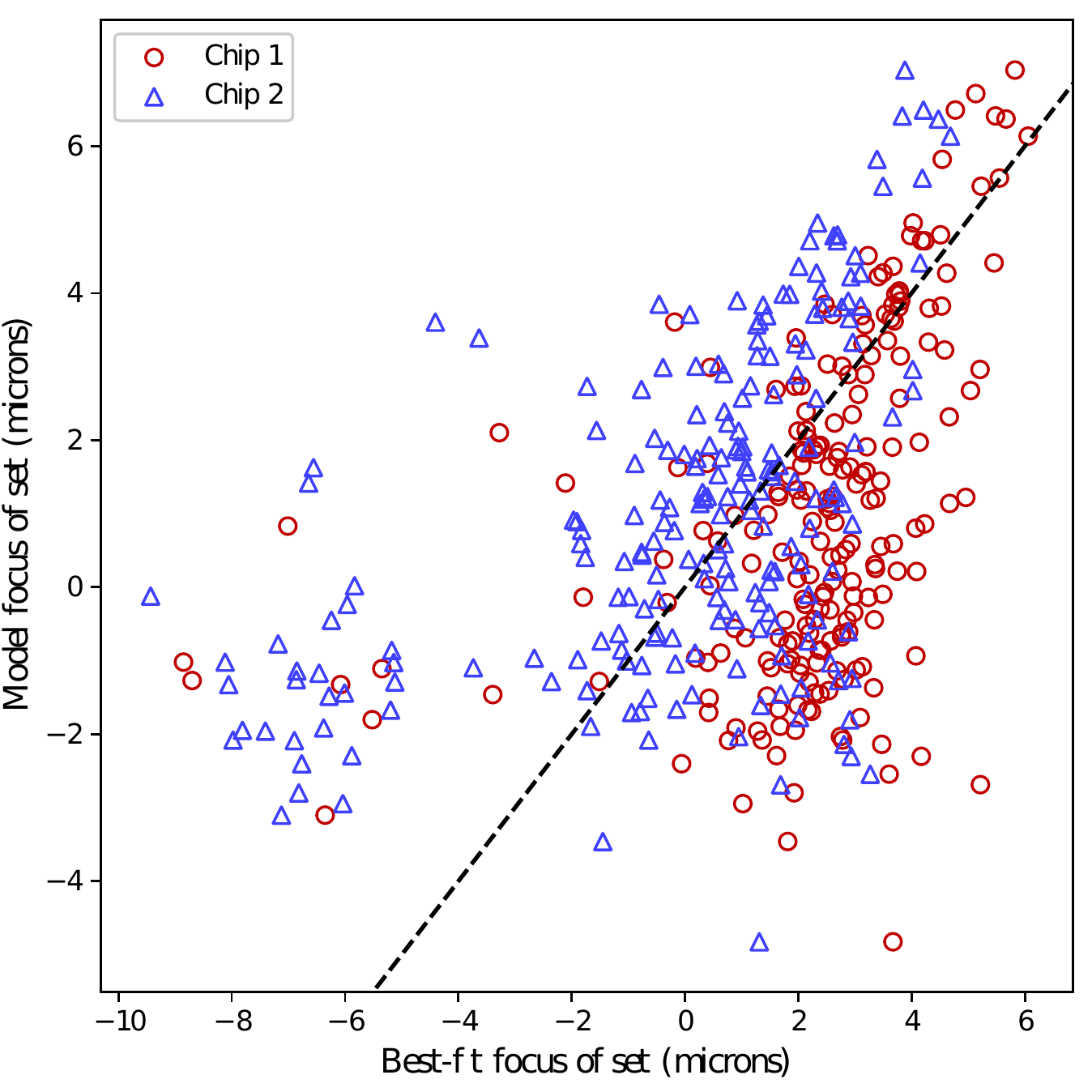}
	\caption{The focus offset, as predicted by the model from \citet{diMakLal08} and \citet{NieLal10}, plotted against our fitted focus offset values from our all-parameter fit. The dashed line indicates an ideal match between the model and our fitted values. Recall that our fitted values are biased away from a focus offset value of $\sim -3$, so the actual focus offset values are likely closer to $\sim -3$ microns than plotted here.}
	\Figlabel{model_focus_v_focus}
\end{figure}

A model to predict the focus offset of HST was developed by \citet{diMakLal08} and \citet{NieLal10}. As reported in \citet{CoxNie11}, from tests comparing the model's predictions to measurements of the focus offset made through the phase-retrieval process described in \citet{KriBur95}, this model predicts the focus offset to within 1 micron $\sim 50$ per cent of the time, and to within 2 microns $\sim 80$ per cent of the time.

In \Figref{model_focus_v_focus} we show a comparison of our fitted focus offset values to those predicted from the model (taken from the Annual Summary files published online, for the time of the exposure). We can see that the fitted and model values correlate well for large values of the focus, but the relationship worsens for lower values. Part of this is due to the known effect in our fitting procedure that focus offset values are biased away from $\sim -3$ microns. However, this effect is not sufficiently large to explain the cluster of images for which our fitted focus is $\sim -7$ while the model focus is $\sim -2$.

We confirm the finding of \citet{CoxNie11} that there is a significant difference between the two chips in comparison to the model, of approximately $0.5$ microns. However, \citet{CoxNie11} attributed this to possible differences in charge diffusion and spherical aberration between the two chips, but our results here fit for these parameters individually for each chip, and this difference remains. This suggests that this is not the full explanation for this effect, and other possibilities will have to be investigated.

Since the model's predictions were validated against measurements to be generally accurate within 2 microns, it is worth considering why our measurements differ by more than this threshold. The key difference between our fitting procedure and the phase-retrieval process used for validation of the model \citep[see]{KriBur95} is that the latter uses only out-of-focus observations to fit the focus offset, while we use in-focus observations. Inaccuracy in the extrapolation from out-of-focus to in-focus measurements, for instance due to unknown non-linearity in the motion of the secondary mirror, could result in these measurements differing. Another notable difference is that the measurements used for validation of the model were all made from images of a star at the centre of the detector, while our measurements use stars from various positions on the detectors. If the model fails to properly account for spatial variation of the PSF (see our discussion of this possibility in \Secref{Init_Testing_Results}), this could also account for the difference.

\subsection{Discussion}
\Seclabel{Discussion}

In \Secref{Init_Testing_Results}, we showed that our PSF validation framework indicates a significant discrepancy between the model PSFs generated by the Tiny Tim tool and the PSFs of observed stars when using the default parameters and only fitting the focus offset. The magnitude of this discrepancy is too large to be explained by the various effects we expect to see in observations but which were not accounted for in the Tiny Tim models (see \Secref{Control_Image_Design}), the most notable of which is the presence of unresolved binary stars in the sample. One possible explanation of this that we highlight is that the model might not fully capture that spatially-varying nature of the PSF, as we see indications that this effect might vary between images, perhaps due to time-dependent temperature gradients distorting the image plane.

We showed in \Secref{Further_TT_Testing} that if the user takes advantage of the advanced configuration options for Tiny Tim and fits all possible optical parameters for each image, the fit can be significantly improved. The improved models unfortunately still fail to match the observed PSFs, but the remaining difference is small enough that it is unlikely to cause issue for most use cases on HST data. It does fail to meet the more stringent requirements for the Euclid mission on additive bias, but it as yet uncertain whether this requirement will in fact need to be met in order for HST images to be suitable for validation of the Euclid pipeline.

Unfortunately, performing a fit on all possible optical parameters is significantly more time-consuming than fitting the focus offset alone, and may not be possible with images which only contain a few stars. We investigated two possible ways to improve the fit without having to fit all parameters, by fitting just the focus offset and using either the mean values from the all-parameters fit or a linear relationship between them and the focus offset. Both of these methods provided notable improvement over the focus-offset-only fit. They did not provide quite as significant a benefit as the all-parameter fit, but they require orders of magnitude less computer time to perform, and are thus recommended. In the case of the use of the best-fit parameters, we calculate that the remaining inaccuracies in the PSF model will cause approximate multiplicative shear-measurement bias of $m\sbr{1}\approx m\sbr{2}\approx 4.6 \times 10^{-4}$ and additive bias of $c\sbr{1}\approx 3.6 \times 10^{-4}$ and $c\sbr{2}\approx 3.0 \times 10^{-4}$ (see the calculations in \Appref{shear_bias_relation} for further details on how this is determined).

In order to aid others, we have provided a Python wrapper script in \Appref{gen_PSF} to generate Tiny Tim PSFs using either the best-fit optical parameters or a linear fit with the focus offset.

In \Secref{qof_parameters}, we had to make an arbitrary choice about what single parameter to minimise in our fitting procedure, and we decided to use $\Chisq$, which is an equally-weighted sum of the $Z^2_k$ values which contribute to additive and multiplicative bias, as we did not beforehand know which would be more of an issue. Now that we have our results, we can see that while the $Z^{2(+)}\sbr{s}$ value was the most significant contributor to $\Chisq$ and had the greatest differences between the model and observed PSFs, it still ended up falling well within the range that would confidently meet the Euclid mission's requirements on multiplicative bias. On the other hand, while the $Z^{2(\pm)}\sbr{+}$ and $Z^{2(\pm)}\sbr{\times}$ values did not contribute as much to $\Chisq$ nor differ as much between the model and observed PSFs, they greatly exceeded the range of values that would be consistent with the requirements on additive bias for the Euclid mission. This suggests that for future work, it would be more useful to weight the $Z^{2(\pm)}\sbr{+}$ and $Z^{2(\pm)}\sbr{\times}$ values more than the $Z^{2(+)}\sbr{s}$ value, so that the contribution to additive bias might be reduced, while the contribution to multiplicative bias is still within requirements.

In \Secref{Model_Comparison}, we compared our fitted focus offset values to those predicted by the model from \citet{diMakLal08} and \citet{NieLal10}. For positive focus offsets, we found a reasonable correspondence between our fitted values and the model's values, and we confirmed the offset they found between the two chips of the detector. However, for negative values of the focus offset, we fitted significantly lower values than were predicted by the model. This suggests that there is some issue either with the Tiny Tim PSFs models for negative focus offsets or with the model's predicted values in this regime. Further investigation into this issue is thus warranted.

\section{Summary and Conclusions}
\Seclabel{Conclusions}

In this paper, we have presented a framework for the validation of models for space-based PSFs, with a particular focus on the requirements for weak gravitational lensing measurements. As lensing measurements are most directly affected by the dipole and quadrupole moments and size of the PSF, we focused on testing these aspects. We developed a framework which tests these parameters, providing indications when a model fails and allowing it to be determined if a failure will be significant for weak gravitational lensing measurements.

As an example of how this framework can be applied in practice, we used it to assess the quality of model PSFs generated by the Tiny Tim tool. To form a basis for analysis, we generated control images on which we first applied the framework. Mock stars were placed on these control images, using PSFs generated from the Tiny Tim tool with a known configuration for each image. We then applied the testing framework to each image and compiled the results, determining the typical statistics we expect from this framework in ideal and more realistic usage scenarios.

We then applied the framework to test model PSFs generated with Tiny Tim against stars observed in HST ACS images. In the first test, we used the default configuration for Tiny Tim, only fitting the focus-secondary-mirror despace for each image. Here we found that there was a significant difference between the models and observations, most notably with the measured sizes of the model PSFs differing from those of the observed PSFs with significantly more scatter than would be predicted from noise alone. This finding of a significant difference between the model and observations is consistent with the findings of \citet{vanBelH12}, who found a similar effect on a smaller sample of stars observed with the F160W filter, and with the findings of \citet{RhoMasAlb07}.

We then tested more advanced usage of Tiny Tim, to see if it is possible for the end-user to generate improved models in some manner. To do this, we tried additionally fitting all Zernike polynomials for the optics up to $Z\sbr{22}$ for each image. In this case, the quality of the fit improved significantly. Although the mismatch between the model and observed PSFs is still statistically significant, the practical significance is minimal for the purposes of weak lensing. Most of this improvement can be retained through the use of the best-fit optical parameters from this fit and only fitting the focus offset, or else by using a linear relationship between these parameters and the focus offset, with a significantly lower computer time cost.

We thus conclude that our testing framework is a valuable tool for assessing the quality of model PSFs for the purposes of weak gravitational lensing. Additionally, through testing the framework on the Tiny Tim tool, we have been able to identify deficiencies with it in its standard configuration and are able to provide a script which will generate improved models, which we supply in \Appref{gen_PSF}.

\section*{Data availability}

The data underlying this article were accessed from the Barbara A. Mikulski Archive for Space Telescopes (\texttt{http://archive.stsci.edu/}). The derived data generated in this research will be shared on reasonable request to the corresponding author.

\section*{Acknowledgments}

The authors acknowledge useful input on the paper from Lance Miller.

The authors acknowledge useful conversations with John Krist about the Tiny Tim tool, Matt Lallo about the secondary mirror of HST and the Tiny Tim PSF models, and Sami Niemi about the HST focus model.

The authors thank the anonymous referee for their input on the paper, including their recommendation to address the issue of jitter and some valuable insight on the issue of the observed bias toward too-large fitted focus offset values.

This work is based on observations made with the NASA/ESA {\it Hubble Space Telescope}, using imaging data from the GO programs 11397, 11586, 11653, 11677, 11688, 11724, 11880, 11887, 12166, 12193, 12385, 12389, 12730, and 12734, obtained via the data archive at the Space Telescope Science Institute.

BRG and ANT thank the UK Space Agency for funding.

TS \& RMas acknowledge support through ERC H2020-COMPET-2017, project \#776247.

OM acknowledges support from the German Federal Ministry for Economic Affairs and Energy (BMWi) provided via DLR under project nos. 50QE1103 and 50QE2002.

RMan was supported by NASA ROSES grant 12-EUCLID12-0004.

RMas is supported by UK Space Agency grants ST/N001494/1 and ST/V001582/1 and a Royal Society University Research Fellowship.

JR was supported by JPL, which is run under a contract for NASA by Caltech and by NASA ROSES grant 12-EUCLID12-0004.

ANT thanks the Royal Society for a Wolfson Research Merit Award.

This project has received funding from the European Union’s Horizon 2020 research and innovation programme under grant agreement No 776247.

\bibliographystyle{mn2e}
\setlength{\bibhang}{2.0em}
\setlength\labelwidth{0.0em}
\bibliography{../Magnification_Method/full_bib}

\appendix

\section{Size Estimator Tests}
\Applabel{Qsize}

In \Secref{relevant_quantities} we explained our need for a background-independent size estimator. In this appendix, we present tests and analysis of this estimator.

\begin{figure*}
	\includegraphics[scale=0.54]{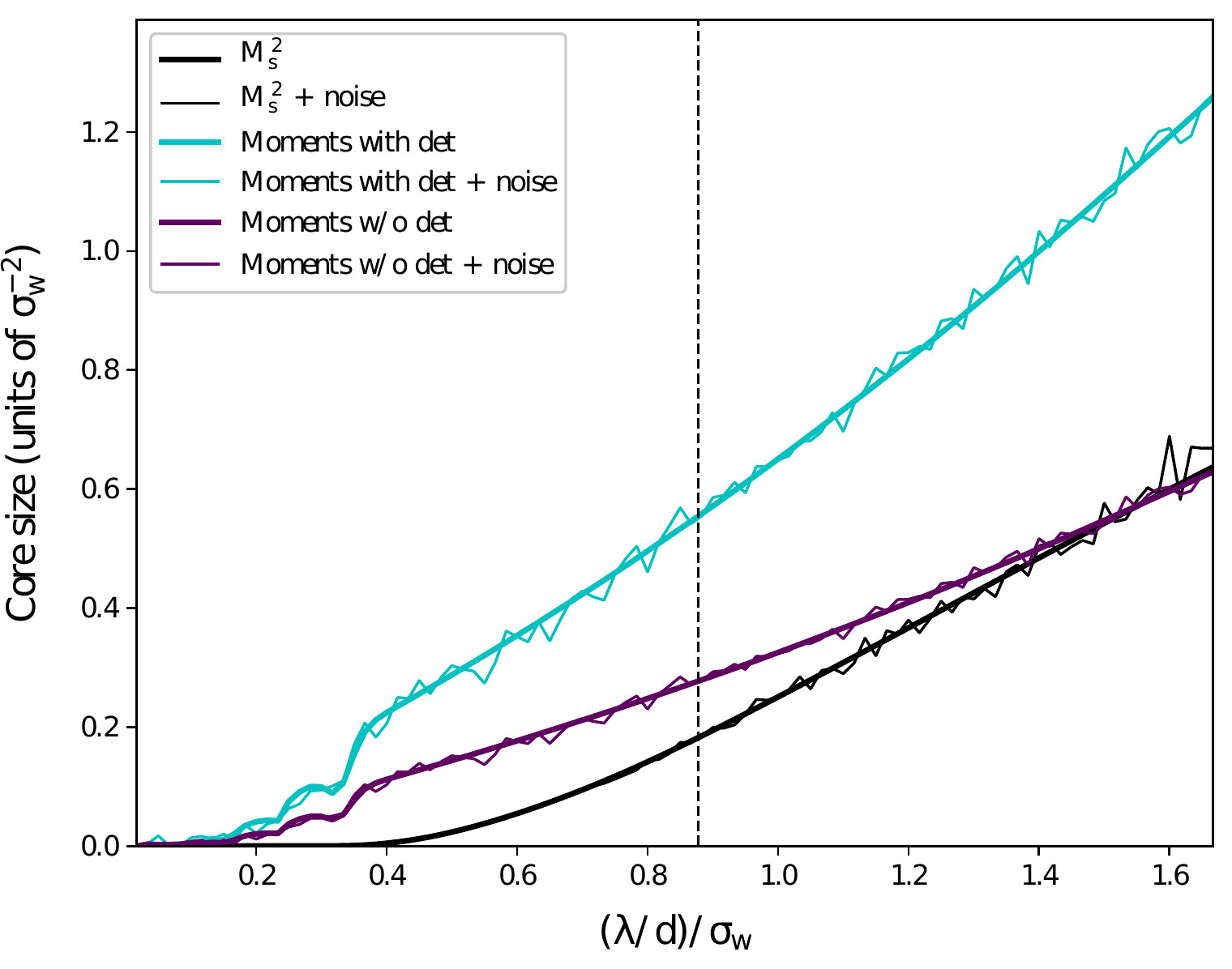}
	\includegraphics[scale=0.54]{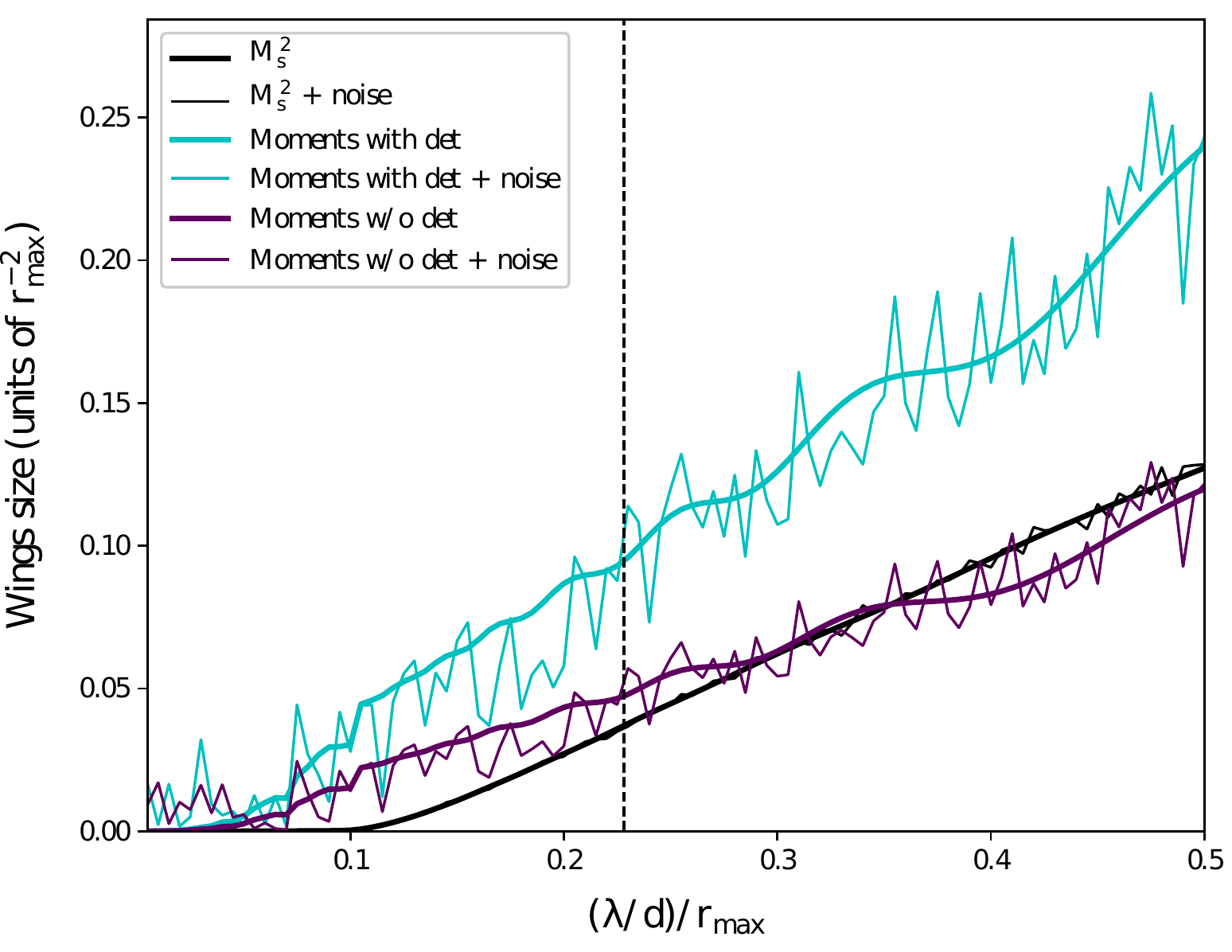}
	\caption[Comparison of $M\sbr{s}$ size estimator with moments-based size estimators.]{Comparison of our background-independent size estimator, $M\sbr{s}$, with moments-based size estimators through tests on an Airy profile of varying size $\lambda/d$, using a weighting function favouring the core of the profile (left) and the wings of the profile (right). $M\sbr{s}$ is squared so that it can be more easily compared to the other estimators. The input and measured radii are scaled by the $\sigma$ of the Gaussian core weight function in the left panel, and by the maximum radius of the tophat wings weight function in the right panel. The pixel scale used for this test is equivalent to $\sigma\sbr{w}/3$ and $r\sbr{max}/10$. The dashed line indicates the typical size of stars in our samples, determined from measuring the size of a stack of all stars used.}
	\Figlabel{size_estimator_comparison}
\end{figure*}

In \Figref{size_estimator_comparison} we show comparisons of $M\sbr{s}$ with two moments-based size estimators: an estimator not including a determinant term, calculated as $M_{xx} + M_{yy}$, and an estimator including a determinant term, calculated as $M_{xx} + M_{yy} + 2\sqrt{\left|\mathbf{M}\right|}$. Both of these estimators have been used for shear-measurement; the former by e.g. \citet{KaiSquBro95}, and the latter by e.g. \citet{SeiSch97}. The comparison illustrates that $M\sbr{s}$, like the other two estimators, is monotonic increasing with the size of the measured profile. It is less sensitive when the size is smaller than one pixel, but it is much less prone to oscillations when the wings weighting function is used.

The comparison also illustrates the response of each estimator to mock noise, applied with the same prescription used in \Secref{PSF_models}, using the same total flux for each profile. This shows that $M\sbr{s}$ has significantly different noise properties from the other estimators. Notably, it has a higher signal-to-noise ratio when the wings weighting function is used.

\begin{figure*}
	\includegraphics[scale=0.54]{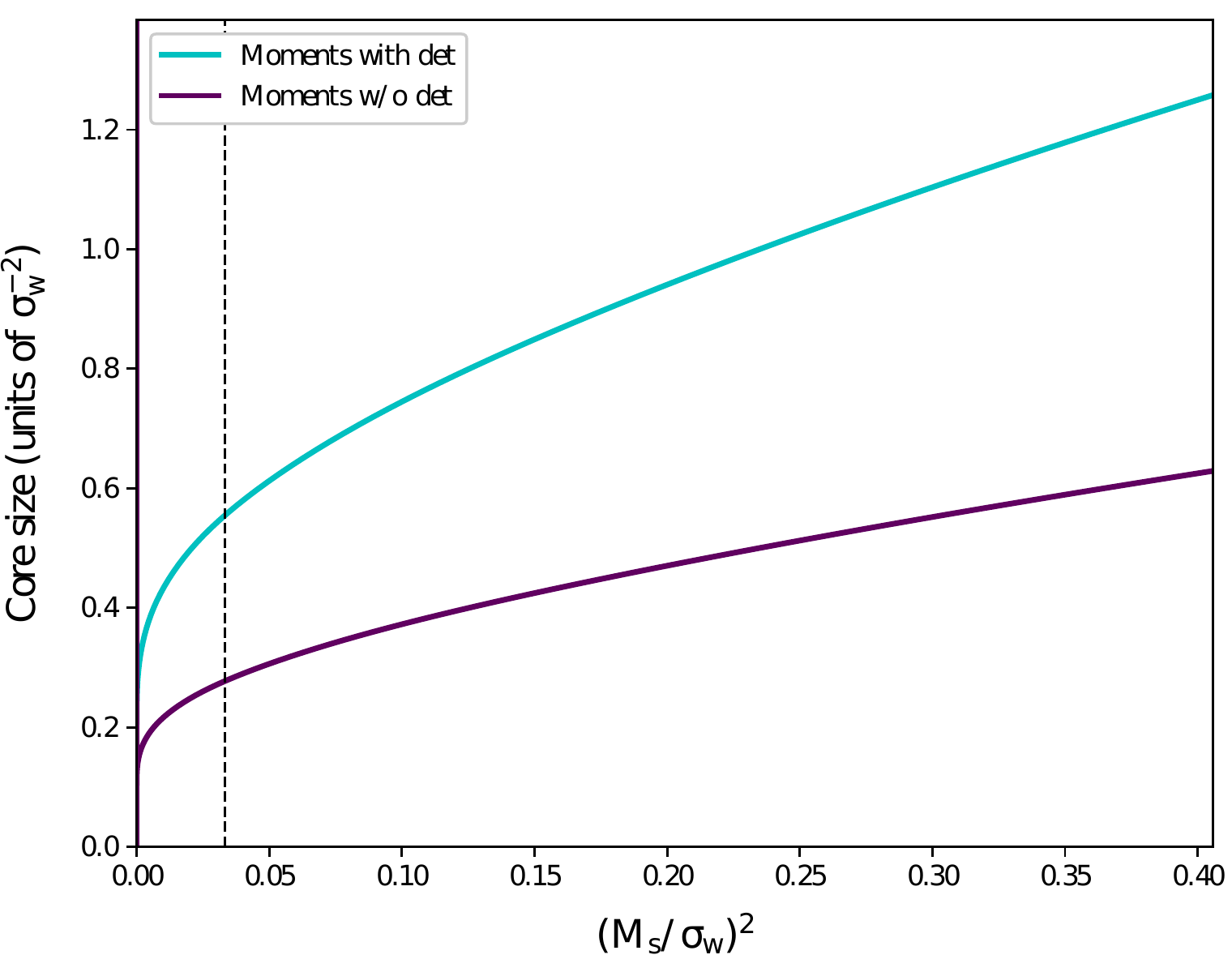}
	\includegraphics[scale=0.54]{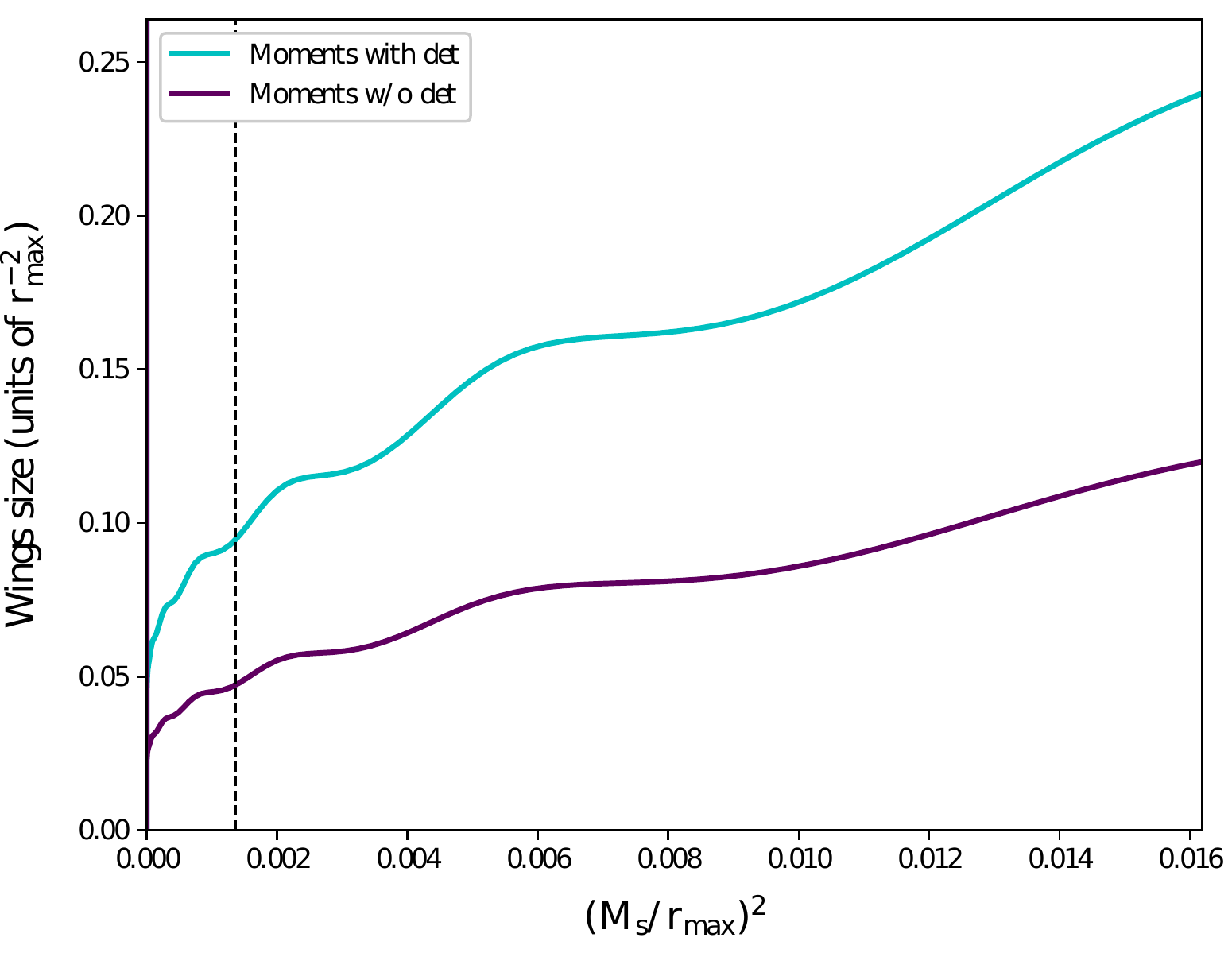}
	\caption[Direct comparison of $M\sbr{s}$ estimator with other estimators.]{As \Figref{size_estimator_comparison}, except with the $Q\sbr{s}$ size estimate on the horizontal axis and the squares of the other two estimates on the vertical axis, and without displaying versions containing mock noise. This illustrates the monotonic increasing relationship between the estimators and allows a conversion of measured $M\sbr{s}$ values to the combinations of moments used in shear measurements.}
	\Figlabel{size_estimator_direct_comparison}
\end{figure*}

It is also useful to directly compare the moment combinations used for ellipticity estimation to the corresponding $M\sbr{s}$ values. This allows us to convert observed differences in $M\sbr{s}$ between models and observed stars to the equivalent changes in these moment combinations, which will allow estimates of how significantly weak lensing measurements would be affected. We show this in \Figref{size_estimator_direct_comparison}, plotting these moment combinations against $M\sbr{s}$.

\section{Relating Fitting Statistics to Shear Bias}
\Applabel{shear_bias_relation}

Using our definition of the relevant moments for shear estimation in \Eqref{relevant_quadrupole_terms}, as well as equations (\ref{eq:ellipticity_from_moments}) and (\ref{eq:moment_subtraction}), we can express the calculation of the ellipticity parameters of the undistorted galaxy as
\begin{align}
	\Eqlabel{ellipticity_from_relevant_moments}
	\hat{e}\sbr{1} &= \frac{M\sbr{+}^{(u)}}{M\sbr{*}^{(u)}} = \frac{M\sbr{+}^{(o)} - M\sbr{+}^{(p)}}{M\sbr{*}^{(o)} - M\sbr{*}^{(p)}}\mathrm{,} \\ \nonumber
	\hat{e}\sbr{2} &= \frac{M\sbr{+}^{(u)}}{M\sbr{*}^{(u)}} = \frac{M\sbr{\times}^{(o)} - M\sbr{\times}^{(p)}}{M\sbr{*}^{(o)} - M\sbr{*}^{(p)}}\mathrm{.} 
\end{align}
Let us now consider how these are affected by perturbations in $\mathbf{M}^{(p)}$, as might be caused by an imperfect PSF model being used.

Starting with $M\sbr{+}$:
\begin{align}
	\hat{e}\sbr{1}(1+m\sbr{1})+c\sbr{1} &= \frac{M\sbr{+}^{(o)} - \left( M\sbr{+}^{(p)} + \delta M\sbr{+}^{(p)} \right) }{M\sbr{*}^{(o)} - M\sbr{*}^{(p)}} \\ \nonumber
		&= \frac{M\sbr{+}^{(o)} - M\sbr{+}^{(p)}}{M\sbr{*}^{(o)} - M\sbr{*}^{(p)}} - \frac{\delta M\sbr{+}^{(p)}}{M\sbr{*}^{(o)} - M\sbr{*}^{(p)}} \mathrm{,}
\end{align}
while $\hat{e}\sbr{2}$ is unchanged. This implies that a perturbation in the $M\sbr{+}$ of the PSF model will result in an additive bias on $\hat{e}\sbr{1}$, with the relationship
\begin{equation}
	c\sbr{1}  = -\frac{1}{M\sbr{*}^{(u)}}\delta M\sbr{+}^{(p)} \mathrm{.}
\end{equation}
Similarly, we can calculate
\begin{equation}
	c\sbr{2}  = -\frac{1}{M\sbr{*}^{(u)}}\delta M\sbr{\times}^{(p)} \mathrm{.}
\end{equation}
For $M\sbr{*}$:
\begin{align}
	\hat{e}\sbr{1}(1+m\sbr{1})+c\sbr{1} &= \frac{M\sbr{+}^{(o)} - M\sbr{+}^{(p)} }{M\sbr{*}^{(o)} - \left( M\sbr{*}^{(p)}  + \delta M\sbr{*}^{(p)}\right)} \\ \nonumber
		&\approx \frac{M\sbr{+}^{(o)} - M\sbr{+}^{(p)}}{M\sbr{*}^{(o)} - M\sbr{*}^{(p)}} \left(1 + \frac{\delta M\sbr{*}^{(p)}}{M\sbr{*}^{(o)} - M\sbr{*}^{(p)}} \right) \mathrm{,}
\end{align}
and similarly for $\hat{e}\sbr{2}$, so
\begin{equation}
	m\sbr{1} = m\sbr{2}  = \frac{1}{M\sbr{*}^{(u)}}\delta M\sbr{*}^{(p)} \mathrm{.}
\end{equation}

$M_{x}$ and $M_{y}$ do not directly enter the equations for estimating ellipticity, but are used instead to determine the centres of objects. For a moments-based method, this should generally not impact the ellipticity estimate as long as the PSF is centred using the same weight function as the galaxy. For model-fitting methods, however, we cannot rule out that this might have an impact. To estimate the maximum magnitude of this impact, we will consider the worst-case scenario for a moments-based method, where one of our two extreme weighting functions is used to centre the PSF, and the other is used to measure its moments.

If the difference in moments between the two weight functions is $\delta(M_{x}^c - M_{x}^w)$ when measured near the centre, and this is small relative to the scale of the weight functions, then $\delta(M_{x}^c - M_{x}^w)$ will approximate the difference in centre positions for the two weight functions. It can be shown from \Eqref{moment_definitions} that when measured offset from the centre by a small distance $d_x$, $M_{xx}$ will increase by $d_x^2$. The situation is similar for $M_{yy}$. This implies that a difference in these moments between the weight function can possibly result in changes to the second-order moments, contributing both additive and multiplicative bias of magnitudes
\begin{align}
	|m| &\approx \frac{4}{M\sbr{*}^{(u)}}\left[\left(\delta(M_{x}^c - M_{x}^w)\right)^2+\left(\delta(M_{y}^c - M_{y}^w)\right)^2\right] \mathrm{,} \\ \nonumber
	|c| &\approx \frac{1}{M\sbr{*}^{(u)}}\left[\left(\delta(M_{x}^c - M_{x}^w)\right)^2+\left(\delta(M_{y}^c - M_{y}^w)\right)^2\right] \mathrm{.}
\end{align}
The factor $4$ in the $m$ bias is due to a factor of $2$ from $M_{xx}$ and $M_{yy}$ each showing up twice in the definition of $M_{*}$ and another factor of $2$ since any change in the size affects both $m\sbr{1}$ and $m\sbr{2}$. 

Since we wish to combine all statistics together for a single quality-of-fit parameter, we must decide how to weight additive versus multiplicative bias. For the sake of simplicity, we decide to weight them both equally for this work. We thus have the following scaling relationships between shear bias $b = m\sbr{1} + m\sbr{2} + c\sbr{1} + c\sbr{2}$ and each parameter:
\begin{align}
	\Eqlabel{db_dM}
	\frac{\delta b}{\delta M_{x}^{(p)2} } = \frac{5}{M\sbr{*}^{(u)}} \mathrm{,} \\ \nonumber
	\frac{\delta b}{\delta M_{y}^{(p)2} } = \frac{5}{M\sbr{*}^{(u)}} \mathrm{,} \\ \nonumber
	\frac{\delta b}{\delta M\sbr{+}^{(p)} } = \frac{1}{M\sbr{*}^{(u)}} \mathrm{,} \\ \nonumber
	\frac{\delta b}{\delta M\sbr{\times}^{(p)} } = \frac{1}{M\sbr{*}^{(u)}} \mathrm{,} \\ \nonumber
	\frac{\delta b}{\delta M\sbr{*}^{(p)} } = \frac{2}{M\sbr{*}^{(u)}} \mathrm{.}
\end{align}
with the proviso that the bias relationships for $M_{x}^{(p)}$ and $M_{x}^{(p)}$ are estimates of a worst-case scenario and only relevant for the difference in measurements between weight functions.

We can use these expressions if we wish to relate a perturbation in $\mathbf{M}^{(p)}$ to the amount of bias which will result if we know the $M\sbr{*}^{(u)}$ of the undistorted galaxy. In order to determine estimates representative of the worst-case scenario, we assume that the smallest possible galaxies are being measured, with their size comparable to the size of the PSF. Here, this corresponds to a typical size of $M\sbr{*}^{(u)} \sim 5 \px^2$ when measured with the core weighting function, and $M\sbr{*}^{(u)} \sim 9 \px^2$ with the wings weighting function.

For our fitting procedure, we decided to linearise the relevant parameters for each weight function through \Eqref{Q_definitions}, as well as to use an alternative, background-independent size estimator in place of $M\sbr{*}$. For the size estimator, we can see from \Figref{size_estimator_comparison} that $M\sbr{s}$ has a nearly linear relationship with $M\sbr{*}$ for an Airy profile, with a slope of
\begin{equation}
	\frac{\delta M\sbr{*}^{(p)}}{\delta M\sbr{s}^{(p)}} \approx 1.5 \mathrm{.}
\end{equation}
We can thus calculate
\begin{equation}
	\Eqlabel{db_dMs}
	\frac{\delta b}{\delta M\sbr{s}^{(p)} } = \frac{\delta b}{\delta M\sbr{*}^{(p)} } \frac{\delta M\sbr{*}^{(p)}}{\delta M\sbr{s}^{(p)}} \approx \frac{3}{M\sbr{*}^{(u)}} \mathrm{.}
\end{equation}

Since we decided to linearise the $M^{c/w}$ values for each weight function into linear combinations $Q^{(\pm)}$, we cannot directly apply Equations (\ref{eq:db_dM}) and (\ref{eq:db_dMs}) without undoing the benefits of linearisation. Instead, since in any case we cannot calculate exact bias projections without knowing what weight functions will be used, we decide to use representative values instead. Making use of the fact that in the scenario where $\sigma(M^{\rm (c)}_k)=\sigma(M^{\rm (w)}_k)$, we will have the property $\left(Q^{(+)}_k\right)^2 + \left(Q^{(-)}_k\right)^2 = \left(M^{\rm c}_k\right)^2 + \left(M^{\rm (w)}_k\right)^2$, we can assume that the $Q^{(\pm)}_k$ values will be of similar magnitude to the $M^{c/w}_k$ values, and thus
\begin{align}
	\frac{\delta b}{\delta Q_k} &\approx \frac{\delta b}{\delta M_k} \mathrm{.}
\end{align}
Finally, since we are using a combination of the core and wings weighting function, we must use a representative size $M\sbr{*}^{(u)}$ intermediate the two size calculations, and so we choose to use the average value $M\sbr{*}^{(u)} = 7 \px^2$.

This gives us the ultimate set of representative bias scaling relationships,
\begin{align}
	\frac{\delta b}{\delta \left(Q_{x}^{(-)}\right)^2 } \approx \frac{5}{7 \px^2} \approx 0.71 \px^{-2} \mathrm{,} \\ \nonumber
	\frac{\delta b}{\delta \left(Q_{y}^{(-)}\right)^2 } \approx \frac{5}{7 \px^2} \approx 0.71 \px^{-2} \mathrm{,} \\ \nonumber
	\frac{\delta b}{\delta Q\sbr{+}^{(\pm)} } \approx \frac{1}{7 \px^2} \approx 0.14 \px^{-2} \mathrm{,} \\ \nonumber
	\frac{\delta b}{\delta Q\sbr{\times}^{(\pm)} } \approx \frac{1}{7 \px^2} \approx 0.14 \px^{-2} \mathrm{,} \\ \nonumber
	\frac{\delta b}{\delta Q\sbr{s}^{(\pm)} } \approx \frac{3}{7 \px^2} \approx 0.43 \px^{-2} \mathrm{,}
\end{align}
which we can use to calculate $Z^2$ values for each statistic, each of which will be representative of the square of the shear bias contributed by each parameter:
\begin{align}
	Z_x^{2(-)} &= \frac{(0.71 \px^{-2})^2}{N}\sum_{i=0}^{N}\left(Q_{x,{\rm star},i}^{(-)}-Q_{x,{\rm model},i}^{(-)}\right)^{4} \mathrm{,} \\ \nonumber
	Z_y^{2(-)} &= \frac{(0.71 \px^{-2})^2}{N}\sum_{i=0}^{N}\left(Q_{x,{\rm star},i}^{(-)}-Q_{x,{\rm model},i}^{(-)}\right)^{4} \mathrm{,} \\ \nonumber
	Z\sbr{+}^{2(\pm)} &= \frac{(0.14 \px^{-2})^2}{N}\sum_{i=0}^{N}\left(Q_{{\rm +},{\rm star},i}^{(\pm)}-Q_{{\rm +},{\rm model},i}^{(\pm)}\right)^2 \mathrm{,} \\ \nonumber
	Z\sbr{\times}^{2(\pm)} &= \frac{(0.14 \px^{-2})^2}{N}\sum_{i=0}^{N}\left(Q_{{\rm \times},{\rm star},i}^{(\pm)}-Q_{{\rm \times},{\rm model},i}^{(\pm)}\right)^2 \mathrm{,} \\ \nonumber
	Z\sbr{s}^{2(\pm)} &= \frac{(0.43 \px^{-2})^2}{N}\sum_{i=0}^{N}\left(Q_{{\rm s},{\rm star},i}^{(\pm)}-Q_{{\rm s},{\rm model},i}^{(\pm)}\right)^2 \mathrm{.} \\ \nonumber
\end{align}

Under the assumptions that these quantities are generally independent and that each is as likely to cause a positive or negative bias, we can treat the sum of these values,
\begin{equation}
	\Chisq = \sum_k Z^2_k \mathrm{,}
	\Eqlabel{Chisq_def2}
\end{equation}
as representative of the square of the total shear bias contributed by inaccuracies in the PSF. This provides us with our desired goal of a single representative quantity which we can minimise in the fitting procedure.

\section{Time-Dependence of Best-Fit Optical Parameters}
\Applabel{params_FT}

In \Secref{Further_TT_Testing}, we showed that fitting all possible optical parameters to each image provided a significantly better match between the model PSFs and observed stars than only fitting the focus offset. When the mean best-fit values for each parameter were used, or when linear relationships between these values and the focus offset were used, the improvement over the focus-offset-only case was not nearly as significant. In this appendix, we illustrate other tests we performed on these best-fit values to test if there is evidence for any time-dependent behaviour.

\begin{figure*}
	\includegraphics[scale=0.68]{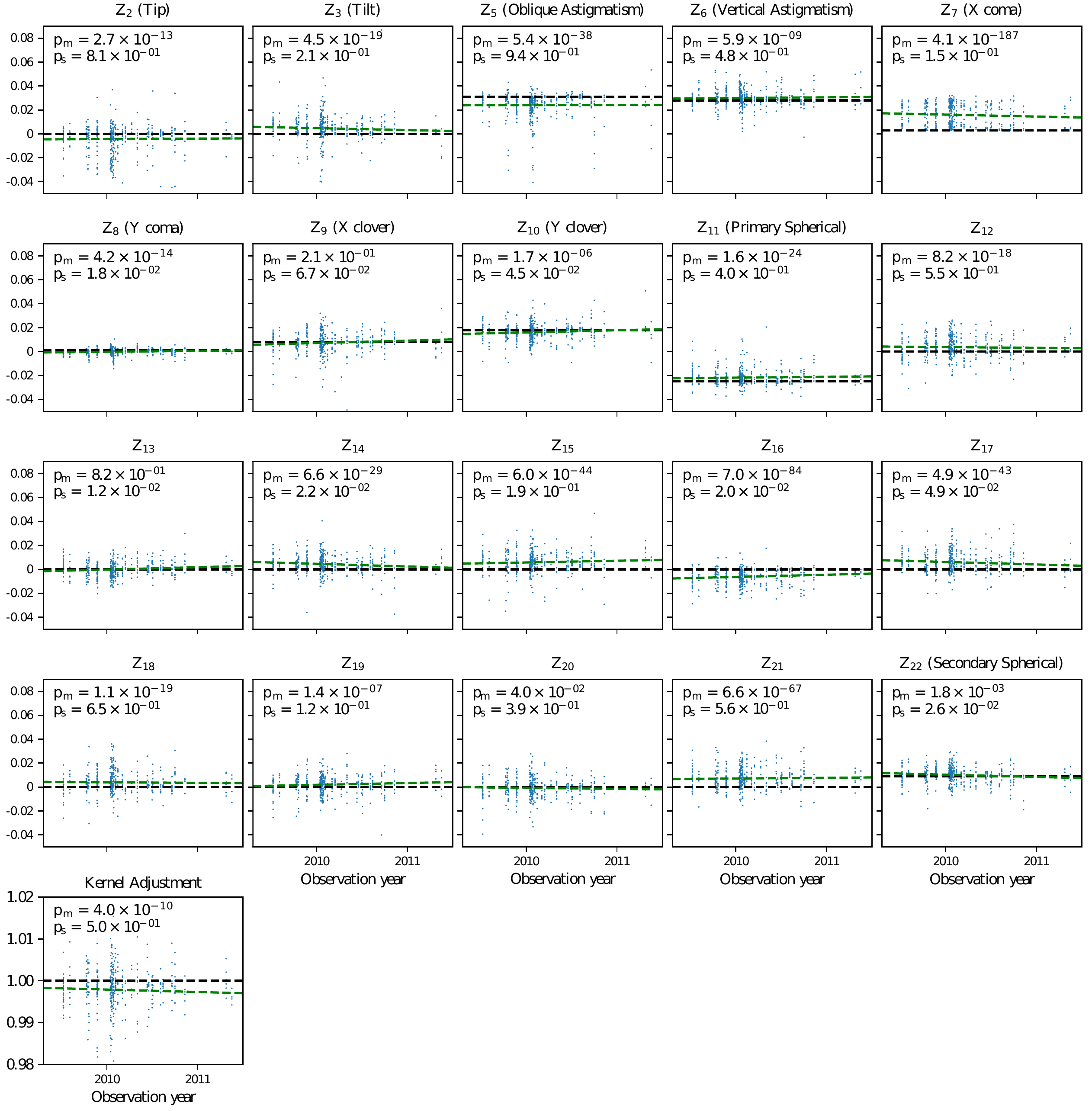}
	\caption[]{The best-fit optical parameters, plus the adjustment $c$ to the charge diffusion kernel, for each image plotted against the date of observation. The solid black line indicates the default value used for each by Tiny Tim, and the dashed blue line is the best linear fit to the data. The p-values stated on each panel are the two-tailed Gaussian probabilities that the mean fitted value is consistent with the default value and that the slope is consistent with zero. When the mean is separated from the default value by at least one standard deviation, or the slope is separated from zero by at least one standard deviation, the respective p-value is emboldened.}
	\Figlabel{bestfit_param_v_obs_time}
\end{figure*}

In \Figref{bestfit_param_v_obs_time}, we plot the best-fit optical parameters and our adjustment to the charge diffusion kernel against the date of observation, and show the best-fit linear relationship. If any parameter varies gradually or discontinuously with time, we would expect to a significantly non-zero slope in the best-fit linear relationship here. However, the slope is nearly zero for all parameters except $Z_{\rm 21}$. With this many parameters tested, it is not unusual that one might have a significantly non-zero slope due to noise alone, and so overall we see no evidence here to indicate a significant gradual or discontinuous relationship between any parameters and the date of observation.

Another possibility is that one or more of the optical parameters might vary periodically with time. To test this, we calculate Fourier modes for each parameter through
\begin{equation}
	A_f = \left| \sum_j (v_j - \left<v_j\right>) \exp{2\pi i f t_j}  \right| \mathrm{,}
\end{equation}
where summation is performed over all images $j$, $v_j$ is the value of a given optical parameter (or the modification to the charge-diffusion kernel) fit for image $j$, $\left<v_j\right>$ is the expected value of $v_j$ from a linear fit to the focus offset, $f$ is the frequency, and $t_j$ is the time of observation for image $j$. The subtraction of $\left<v_j\right>$ here is done to reduce or remove the impact of degeneracies with the focus offset on the calculated amplitudes here. Note that due to sparse sampling, the calculated amplitudes for different frequencies are not necessarily independent.

\begin{figure*}
	\includegraphics[scale=0.45]{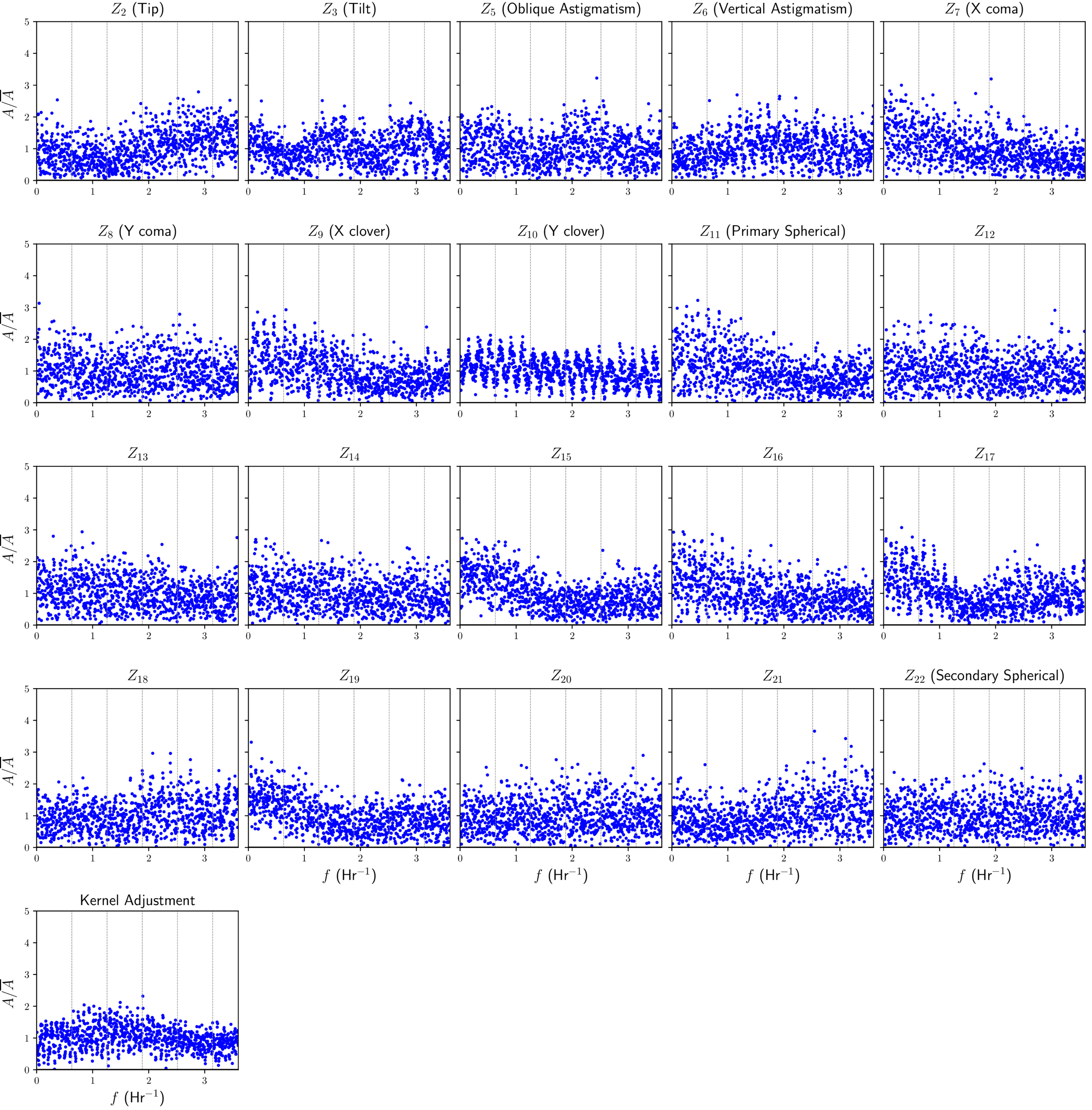}
	\caption[]{The amplitudes of the Fourier modes of all optical parameters when they were all fit to each image. Before calculating Fourier modes, each parameter had a linear fit of it against the focus offset subtracted from its value, so that the known periodic behaviour of the focus offset would not combine with a degeneracy with certain parameters to result in spurious modes appearing. The amplitudes are all normalised by the mean amplitude to highlight notable deviations. The frequency is presented in units of inverse hours. Dashed lines indicate multiples of the HST's orbital frequency.}
	\Figlabel{bestfit_param_FT}
\end{figure*}

We plot the resulting amplitudes in \Figref{bestfit_param_FT}. We would expect any periodic behaviour to manifest as spikes in these plots. Parameters such as $Z_{\rm 7}$ (the ``X coma'') and $Z_{\rm 17}$ show coherent variations in the amplitude across a large range of frequencies, which likely arises simply from coincidental noise in the time-domain values, and does not have any physical meaning.

The dotted lines in this plot correspond to integer multiples of HST's orbital frequency. Any spikes near these values would correspond to variations out-of-phase with changes to the focus, but we see no such spikes. There may however be variations in phase with the focus, but we cannot disentangle these from the effects of degeneracies between the parameters.

Overall, there is not enough evidence to confirm the presence of periodic behaviour in any of the optical parameters, but the data is too noisy to entirely rule it out.

\onecolumn
\section{Generating Tiny Tim PSFs With our Fitted Parameters}
\Applabel{gen_PSF}

\begin{table}
	\caption{The mean values and slope and intercept of a linear fit against focus offset (in microns) for the optical parameters used by Tiny Tim, determined from our all-parameter fits.}
	\Tablabel{fitted_params}
	\begin{center}
		\begin{tabular}{llll}
			Name & Mean & Intercept & Slope \\ \hline 
			$Z_{2}$ (Tip) & $-0.0042$ & $-0.0043$ & $\phs 0.0000$ \\
			$Z_{3}$ (Tilt) & $\phs 0.0046$ & $\phs 0.0048$ & $-0.0002$ \\
			0 degree astigmatism & $\phs 0.0241$ & $\phs 0.0246$ & $-0.0005$ \\
			45 degree astigmatism & $\phs 0.0300$ & $\phs 0.0302$ & $-0.0002$ \\
			X coma & $\phs 0.0159$ & $\phs 0.0172$ & $-0.0012$ \\
			Y coma & $\phs 0.0000$ & $\phs 0.0000$ & $\phs 0.0000$ \\
			X clover & $\phs 0.0074$ & $\phs 0.0079$ & $-0.0004$ \\
			Y clover & $\phs 0.0163$ & $\phs 0.0169$ & $-0.0005$ \\
			Spherical 3rd & $-0.0217$ & $-0.0207$ & $-0.0008$ \\
			$Z_{12}$ & $\phs 0.0037$ & $\phs 0.0002$ & $\phs 0.0005$ \\
			$Z_{13}$ & $\phs 0.0001$ & $\phs 0.0003$ & $-0.0002$ \\
			$Z_{14}$ & $\phs 0.0043$ & $\phs 0.0039$ & $\phs 0.0003$ \\
			$Z_{15}$ & $\phs 0.0059$ & $\phs 0.0061$ & $-0.0002$ \\
			$Z_{16}$ & $-0.0061$ & $-0.0069$ & $\phs 0.0007$ \\
			$Z_{17}$ & $\phs 0.0059$ & $\phs 0.0059$ & $\phs 0.0000$ \\
			$Z_{18}$ & $\phs 0.0039$ & $\phs 0.0034$ & $-0.0004$ \\
			$Z_{19}$ & $\phs 0.0020$ & $\phs 0.0026$ & $-0.0005$ \\
			$Z_{20}$ & $-0.0008$ & $-0.0014$ & $\phs 0.0005$ \\
			$Z_{21}$ & $\phs 0.0072$ & $\phs 0.0062$ & $\phs 0.0008$ \\
			Spherical 5th & $\phs 0.0101$ & $\phs 0.0088$ & $\phs 0.0011$ \\
			Kernel adjustment & $\phs 0.9978$ & $\phs 0.9978$ & $\phs 0.0000$
		\end{tabular}
	\end{center}
\end{table}

In order to aid others in easily generating Tiny Tim PSFs using the best-fit optical parameters we have determined in this paper, we list both the mean values and the parameters for the linear fit in \Tabref{fitted_params}, and we provide Python code which uses them to call Tiny Tim to generate subsampled PSFs. This code is hosted online at \texttt{https://bitbucket.org/brgillis/tinytim\_psfs} and in the electronic version of this document.

\newpage

\begin{lstlisting}
""" This module contains the needed functions to generate a Tiny Tim PSF, using either
    the best-fit optical parameters found by Gillis et al. (2019) or the linear
    relationship with the focus-secondary-mirror despace.
"""

__all__ = ['make_subsampled_model_psf']

import subprocess as sbp
import os

# Default values
default_psf_position = (2048, 1024) # Center of the detector by default
default_focus = -1.0 # Approximately the middle of expected values
default_chip = 1 
default_spec_type = (1, 15) # Use spectrum for a K-type star by default
default_filter_name = 'F606W'
default_detector = 15 # ACS WFC
default_psf_size = 2.0
default_tinytim_path = "../../Program_Files/tinytim-7.5" # Adjust as needed for your own purposes
default_subsampling_factor = 8

# Default optical parameters
optical_params_means = {"z2":                -0.0042,
                        "z3":                 0.0046,
                        "astigmatism_0":      0.0241,
                        "astigmatism_45":     0.0300,
                        "coma_x":             0.0159,
                        "coma_y":             0.0000,
                        "clover_x":           0.0074,
                        "clover_y":           0.0163,
                        "spherical_3rd":     -0.0217,
                        "z12":                0.0037,
                        "z13":                0.0001,
                        "z14":                0.0043,
                        "z15":                0.0059,
                        "z16":               -0.0061,
                        "z17":                0.0059,
                        "z18":                0.0039,
                        "z19":                0.0020,
                        "z20":               -0.0008,
                        "z21":                0.0072,
                        "spherical_5th":      0.0101,
                        "kernel_adjustment":  0.9978}
			
# Linear fit for optical parameters in (intercept, slope)
optical_params_int_and_slopes = {"z2":                (-0.0043,  0.0000),
                                 "z3":                ( 0.0048, -0.0002),
                                 "astigmatism_0":     ( 0.0246, -0.0005),
                                 "astigmatism_45":    ( 0.0302, -0.0002),
                                 "coma_x":            ( 0.0172, -0.0012),
                                 "coma_y":            ( 0.0000,  0.0000), 
                                 "clover_x":          ( 0.0079, -0.0004),
                                 "clover_y":          ( 0.0169, -0.0005),
                                 "spherical_3rd":     (-0.0207, -0.0008),
                                 "z12":               ( 0.0002,  0.0005),
                                 "z13":               ( 0.0003, -0.0002),
                                 "z14":               ( 0.0039,  0.0003),
                                 "z15":               ( 0.0061, -0.0002),
                                 "z16":               (-0.0069,  0.0007),
                                 "z17":               ( 0.0059,  0.0000),
                                 "z18":               ( 0.0034, -0.0004),
                                 "z19":               ( 0.0026, -0.0005),
                                 "z20":               (-0.0014,  0.0005),
                                 "z21":               ( 0.0062,  0.0008),
                                 "spherical_5th":     ( 0.0088,  0.0011),
                                 "kernel_adjustment": ( 0.9978,  0.0000),}


def replace_multiple_in_file(input_filename, output_filename, input_strings, output_strings):
    """ Replaces every occurence of an input_string in input_filename with the corresponding
        output string and prints the results to $output_filename.
        
        @param[in]  input_filename <str>
        @param[out] output_filename <str>
        @param[in]  input_strings <iterable of strs>
        @param[in]  output_strings <iterable of strs>
                  
        @return None
    """
    
    with open(output_filename, "w") as fout:
        with open(input_filename, "r") as fin:
            for line in fin:
                new_line = line
                for input_string, output_string in zip(input_strings, output_strings):
                    if((input_string is None) or (output_string is None)):
                        continue
                    new_line = new_line.replace(input_string, output_string)
                fout.write(new_line)
                
    return

def make_subsampled_model_psf(filename,
                              psf_position = default_psf_position,
                              focus = default_focus,
                              chip = default_chip,
                              spec_type = default_spec_type,
                              detector = default_detector,
                              filter_name = default_filter_name,
                              psf_size=default_psf_size,
                              tinytim_path = default_tinytim_path,
                              subsampling_factor = default_subsampling_factor,
                              linear_fit = False,
                              clobber = True,
                              **optical_params):
    """ Generates a subsampled model PSF, using the desired (or default) optical parameters.
        For input parameters spec_type and detector, the allowed options can be seen through
        running tiny1
    
        @param[out] filename <str> Desired filename of the generated PSF. If it already exists, the
                                   'clobber' parameter will determine whether or not it will be
                                   overwritten.
        @param[in]  psf_position <(float, float)> Position on the detector of the PSF in x, y
        @param[in]  focus <float> Focus-secondary-mirror despace of the PSF
        @param[in]  chip <int> Which chip the model PSF is for. Allowed values are 1 and 2
        @param[in]  spec_type <(int, *)> Spectral type of the PSF to generate. First value chooses type
                                         of spectrum, second chooses from options for this type
        @param[in]  detector <int> Index of detector to be used.
        @param[in]  filter_name <str> Name of the filter to use (eg. F606W)
        @param[in]  psf_size <float> Size of the PSF image in arcseconds
        @param[in]  tinytim_path <str> Location of the Tiny Tim executables
        @param[in]  subsampling_factor <int> Factor by which to subsample the PSF
        @param[in]  linear_fit <bool> If False, unspecified optical parameters will be given values
                                      based on the mean from Gillis et al. (2018)'s testing. If True,
                                      will use the linear fit from the analysis instead
        @param[in]  clobber <bool> Whether or not to overwrite the target file if it already exists.
        @param[in]  optical_params <dict> Optical parameters aside from focus for this PSF. If not
                                      specified here, defaults will be used based on the
                                      linear_fit parameter.

        @return None
                                 

    """
    
    # If clobber is False, check if the desired file already exists
    if not clobber:
        if os.path.isfile(filename):
            raise IOError("File " + filename + " already exists. Set clobber=True if you wish to overwrite it.")
    
    # Create a directory to contain this project
    try:
        os.makedirs(os.path.split(filename)[0])
    except OSError as e:
        if not ("[Errno 17] File exists:" in str(e) or "[Errno 2] No such file or directory: ''" in str(e)):
            raise
        else:
            pass # No need to raise if the directory already exists

    filename_base = filename.replace(".fits", "")
    if not filename_base + ".fits" == filename:
        raise ValueError("Filename (" + filename + ") must end in '.fits'.")
    
    par_file = filename_base + ".par"
    tmp_par_file = filename_base + ".par.tmp"

    # Set up the command to call tiny1 and execute it
    cmd = "export TINYTIM=" + tinytim_path + "\n" + \
          tinytim_path + "/tiny1 " + tmp_par_file + " << EOF \n" + \
          str(detector) + "\n" + \
          str(chip) + "\n" + \
          str(psf_position[0]) + " " + str(psf_position[1]) + "\n" + \
          str(filter_name) + "\n" + \
          str(spec_type[0]) + "\n" + \
          str(spec_type[1]) + "\n" + \
          str(psf_size) + "\n" + \
          str(focus) + "\n" + \
          filename_base + "\nEOF"
    
    sbp.call(cmd, shell=True)
    
    # Determine which optical parameters we'll be using
    optical_params_to_use = {}
    for param in optical_params_means:
        if param in optical_params:
            optical_params_to_use[param] = optical_params[param]
        elif linear_fit:
            intercept, slope = optical_params_int_and_slopes[param]
            optical_params_to_use[param] = intercept + focus*slope
        else:
            optical_params_to_use[param] = optical_params_means[param]
    
    # Edit the parameter file to adjust optical parameters
    strs_to_replace = []
    replacements = []

    strs_to_replace.append("0.       # Z2 = X (V2) tilt")
    replacements.append(str(optical_params_to_use["z2"]) + "       # Z2 = X (V2) tilt")

    strs_to_replace.append("0.       # Z3 = Y (V3) tilt")
    replacements.append(str(optical_params_to_use["z3"]) + "       # Z3 = Y (V3) tilt")

    strs_to_replace.append("0.031    # Z5 = 0 degree astigmatism")
    replacements.append(str(optical_params_to_use["astigmatism_0"]) + "    # Z5 = 0 degree astigmatism")

    strs_to_replace.append("0.028    # Z6 = 45 degree astigmatism")
    replacements.append(str(optical_params_to_use["astigmatism_45"]) + "    # Z6 = 45 degree astigmatism")

    strs_to_replace.append("0.003    # Z7 = X (V2) coma")
    replacements.append(str(optical_params_to_use["coma_x"]) + "    # Z7 = X (V2) coma")

    strs_to_replace.append("0.001    # Z8 = Y (V3) coma")
    replacements.append(str(optical_params_to_use["coma_y"]) + "    # Z8 = Y (V3) coma")

    if chip==1:
        strs_to_replace.append("0.008    # Z9 = X clover")
    else:
        strs_to_replace.append("0.007    # Z9 = X clover")
    replacements.append(str(optical_params_to_use["clover_x"]) + "    # Z9 = X clover")

    strs_to_replace.append("0.018    # Z10 = Y clover")
    replacements.append(str(optical_params_to_use["clover_y"]) + "    # Z10 = Y clover")

    strs_to_replace.append("-0.025    # Z11 = 3rd order spherical")
    replacements.append(str(optical_params_to_use["spherical_3rd"]) + "    # Z11 = 3rd order spherical")

    strs_to_replace.append("0.       # Z12 = 0 degree Spherical astigmatism")
    replacements.append(str(optical_params_to_use["z12"]) + "       # Z12 = 0 degree Spherical astigmatism")

    strs_to_replace.append("0.       # Z13 = 45 degree Spherical astigmatism")
    replacements.append(str(optical_params_to_use["z13"]) + "       # Z13 = 45 degree Spherical astigmatism")

    strs_to_replace.append("0.       # Z14 = X Ashtray")
    replacements.append(str(optical_params_to_use["z14"]) + "       # Z14 = X Ashtray")

    strs_to_replace.append("0.       # Z15 = Y Ashtray")
    replacements.append(str(optical_params_to_use["z15"]) + "       # Z15 = Y Ashtray")

    strs_to_replace.append("0.       # Z16")
    replacements.append(str(optical_params_to_use["z16"]) + "       # Z16")

    strs_to_replace.append("0.       # Z17")
    replacements.append(str(optical_params_to_use["z17"]) + "       # Z17")

    strs_to_replace.append("0.       # Z18")
    replacements.append(str(optical_params_to_use["z18"]) + "       # Z18")

    strs_to_replace.append("0.       # Z19")
    replacements.append(str(optical_params_to_use["z19"]) + "       # Z19")

    strs_to_replace.append("0.       # Z20")
    replacements.append(str(optical_params_to_use["z20"]) + "       # Z20")

    strs_to_replace.append("0.       # Z21")
    replacements.append(str(optical_params_to_use["z21"]) + "       # Z21")

    strs_to_replace.append("0.009    # Z22 = 5th order spherical")
    replacements.append(str(optical_params_to_use["spherical_5th"]) + "    # Z22 = 5th order spherical")

    replace_multiple_in_file(tmp_par_file, par_file, strs_to_replace, replacements)

    # Set up the command to call tiny2
    cmd = "export TINYTIM=" + tinytim_path + "\n" + \
          tinytim_path + "/tiny2 " + par_file
    # Run the command to call tiny2
    sbp.call(cmd, shell=True)

    # Set up the command to call tiny3
    cmd = "export TINYTIM=" + tinytim_path + "\n" + \
          tinytim_path + "/tiny3 " + par_file + " SUB=" + \
          str(int(subsampling_factor))
    # Run the command to call tiny3
    sbp.call(cmd, shell=True)
    
    # PSF should be generated, now move it to the desired filename
    init_filename = filename_base + "00.fits"
    os.rename(init_filename,filename)

    # Clean up unneeded files. Silently suppress any errors here
    try:
        os.remove(filename_base + "00_psf.fits")
    except OSError as _e:
        pass
    try:
        os.remove(filename_base + ".tt3")
    except OSError as _e:
        pass
    try:
        os.remove(init_filename)
    except OSError as _e:
        pass
    try:
        os.remove(par_file)
    except OSError as _e:
        pass
    try:
        os.remove(tmp_par_file)
    except OSError as _e:
        pass

    return
\end{lstlisting}

\label{lastpage}

\end{document}